\documentclass[12pt]{article}

\usepackage{graphicx,epsf,amssymb,amsbsy,amsfonts,amssymb,amsmath}
\usepackage[dvips]{color}
\usepackage{bbm}

\newif\ifdraft
\drafttrue
\newif\ifpreprint
\preprinttrue

\def\Sect#1{Section~{\ref{#1}}}
\def\sect#1{section~{\ref{#1}}}
\def\fig#1{fig.~{\ref{#1}}}
\def\Fig#1{Fig.~{\ref{#1}}}

\def\tab#1{table~{\ref{#1}}}
\def\Tab#1{Table~{\ref{#1}}}

\def\spa#1.#2{\left\langle#1\,#2\right\rangle}
\def\spb#1.#2{\left[#1\,#2\right]}
\def\spash#1.#2{\spa{\smash{#1}}.{\smash{#2}}}
\def\spbsh#1.#2{\spb{\smash{#1}}.{\smash{#2}}}
\def\sand#1.#2.#3{%
\left\langle\smash{#1}{\vphantom1}^{-}\right|{#2}%
\left|\smash{#3}{\vphantom1}^{-}\right\rangle}
\def\sandpp#1.#2.#3{%
\left\langle\smash{#1}{\vphantom1}^{+}\right|{#2}%
\left|\smash{#3}{\vphantom1}^{+}\right\rangle}
\def\sandpm#1.#2.#3{%
\left\langle\smash{#1}{\vphantom1}^{+}\right|{#2}%
\left|\smash{#3}{\vphantom1}^{-}\right\rangle}
\def\sandmp#1.#2.#3{%
\left\langle\smash{#1}{\vphantom1}^{-}\right|{#2}%
\left|\smash{#3}{\vphantom1}^{+}\right\rangle}

\def\twoloop{{2 \mbox{-} \rm loop}}
\def\m{\mu}
\def\n{\nu}
\def\r{\rho}

\def\I{{\cal I}}
\def\del{\partial}
\def\tree{{\rm tree}}
\def\pol{\varepsilon}
\def\Tr{\, {\rm Tr}}

\def\e{\epsilon}

\def\nn{\nonumber}

\def\eqn#1{eq.~(\ref{#1})}
\def\Eqn#1{Equation~(\ref{#1})}
\def\eqns#1#2{eqs.~(\ref{#1}) and~(\ref{#2})}

\newcommand{\CN}{\mathcal{N}}      % supersymmetry counter
\newcommand{\CP}{\mathcal{P}}
\newcommand{\ZZ}{\mathbbm Z}

\def\NeqFour{{{\cal N}=4}}

\def\NeqEight{{{\cal N}=8}}

\def\be{\begin{equation}}
\def\ee{\end{equation}}
\def\bea{\begin{eqnarray}}
\def\eea{\end{eqnarray}}
\def\ba{\begin{eqnarray}}
\def\ea{\end{eqnarray}}

\def\ksl{\s{k}}

\def\lr{\leftrightarrow}

\def\ksl{\s{k}}

\def\tree{{\rm tree}}
\def\oneloop{{\rm 1\hbox{-}loop}}
\def\twoloop{{\rm 2\hbox{-}loop}}

\newbox\charbox
\newbox\slabox
\def\s#1{{      % Feynman slash
        \setbox\charbox=\hbox{$#1$}
        \setbox\slabox=\hbox{$/$}
        \dimen\charbox=\ht\slabox
        \advance\dimen\charbox by -\dp\slabox
        \advance\dimen\charbox by -\ht\charbox
        \advance\dimen\charbox by \dp\charbox
        \divide\dimen\charbox by 2
        \raise-\dimen\charbox\hbox to \wd\charbox{\hss/\hss}
        \llap{$#1$} }}
\def\ksl{\s{k}}
\def\P{{\rm P}}
\def\NP{{\rm NP}}

\def\subtractfour#1{\ifthenelse{#1=5}{1}{\ifthenelse{#1=6}{2}
{\ifthenelse{#1=7}{3}{\ifthenelse{#1=8}{4}{\ifthenelse{#1=9}{5}
{\ifthenelse{#1=10}{6}{\ifthenelse{#1=11}{7}{\ifthenelse{#1=12}{8}
{\ifthenelse{#1=13}{9}{\ifthenelse{#1=14}{10}{}}}}}}}}}}}

\begin{document}

\ifpreprint
\hfill SLAC--PUB--14091
\fi

\vskip 1.0cm 
\centerline{\Large \bf Ultraviolet Behavior 
of \,${\cal N}$\,=\,\,8 Supergravity\footnote{Presented at 
International School of Subnuclear Physics, 
47th Course, Erice Sicily, August 29-September 7, 2009.}}

\vskip .3 cm 
\begin{center}
Lance~J.~Dixon
\end{center}

\begin{center}
SLAC National Accelerator Laboratory\\
Stanford University\\
Stanford, CA 94309, USA
\end{center}

\date{April, 2010}

\begin{abstract}
In these lectures I describe the remarkable ultraviolet
behavior of $\NeqEight$ supergravity, which through four loops
is no worse than that of $\NeqFour$ super-Yang-Mills theory (a finite
theory).  I also explain the computational tools that allow multi-loop
amplitudes to be evaluated in this theory --- the KLT relations and the
unitarity method --- and sketch how ultraviolet divergences are
extracted from the amplitudes.
\end{abstract}

%\maketitle

%%%%%%%%%%%%%%%%%%%%%%%%%%%%%%%%

\section{Introduction}
\label{Introduction}

Quantum gravity is nonrenormalizable by power counting, because the 
coupling, Newton's constant, $G_N = 1/M_{\rm Pl}^2$, is dimensionful.
In contrast, the coupling constants of unbroken gauge theories, such as
$\alpha$ for QED and $\alpha_s$ for QCD, are dimensionless.
At each loop order in perturbation theory, ultraviolet divergences
in quantum gravity should get worse and worse, in comparison with
gauge theory: There are two more powers of loop momentum in each
loop, to compensate dimensionally for the two powers of the Planck
mass in the denominator of the gravitational coupling.  Instead
of the logarithmic divergences of gauge theory, which are renormalizable
using a finite set of counterterms, quantum gravity should possess
an infinite set of counterterms.
 
String theory is well known to cure the divergences of quantum gravity.
It does so by introducing a new length scale, related to the string tension, 
at which particles are no longer pointlike.  The question these
lectures will address is:  Is this necessary?   Or could 
enough symmetry allow a point-particle theory of quantum gravity 
to be perturbatively finite in the ultraviolet?  If the latter is true, 
even if in a ``toy model'', it would have a big impact on how we 
think about quantum gravity.   While string theory makes quantum 
gravity finite, it does so at the price of having a huge number of
ground state vacua, perhaps $10^{500}$.  Is it possible that there
are consistent theories of quantum gravity with fewer degrees of
freedom, and fewer ground states?

The particular approach we will take in these lectures is to see
whether the divergences generic to point-like theories of 
quantum gravity can be cured using (in part) a large amount of symmetry,
particularly supersymmetry.  The maximal supersymmetry available for a
theory with a maximal spin of two is $\NeqEight$.  In $\NeqEight$ supergravity,
eight applications of the spin-$1/2$ supercharges $Q_i$, $i=1,2,\ldots,8$,
connect the helicity $h=+2$ graviton state with its CPT conjugate state
with helicity $h=2-8\times1/2 = -2$.  Each anti-commuting supercharge can 
be applied either once or not at all, so
the total number of massless states is $2^8=256$.
The complete Lagrangian for this theory was obtained by Cremmer and
Julia~\cite{CremmerJulia} in the late 1970s, following earlier work
by de Wit and Freedman~\cite{deWitFreedman} and by Cremmer, Julia and
Scherk~\cite{CremmerJuliaScherk}.  While this theory has maximal
supersymmetry, it seems unlikely that supersymmetry alone can render it
finite to all orders in perturbation theory; other symmetries or
dynamical principles may have to come in to play.

Many other approaches to quantum gravity have been considered. For example,
the asymptotic safety program~\cite{AsymptoticSafety} posits that there is
a nontrivial ultraviolet fixed point in the exact renormalization group.
Ho\v{r}ava~\cite{Horava} has also proposed a renormalization group flow
solution, in which the ultraviolet fixed point is not Lorentz symmetric;
space and time scale differently at the fixed point.
These proposals are very interesting, but usually require truncation
of the number of operators or other assumptions.  In contrast, here we
will work in a conventional perturbative framework, with an action
that is quadratic in derivatives.

The perturbative ultraviolet behavior of gravity theories in general,
and $\NeqEight$ supergravity in particular, has been under investigation
for a few decades.  Through about 1998, the general suspicion was that
$\NeqEight$ supergravity in four space-time dimensions would first
diverge at three
loops, based on the existence of an $\NeqEight$ supersymmetric local
counterterm at this order~\cite{DeserKayStelle,Ferrara1977mv,Deser1978br,%
Howe1980th,Kallosh1980fi}.  However, in 1998 the two-loop four-graviton
amplitude was computed~\cite{BDDPR}, and its ultraviolet behavior was
found to be better than expected, leading to the speculation that the
first divergence might occur at five loops.  In 2002, Howe and Stelle
speculated~\cite{HoweStelleNew}, based on the possible existence of a 
superspace formalism realizing seven of the eight supersymmetries, 
that the divergence might be delayed until six loops.  However,
a more recent analysis by Bossard, Howe and Stelle~\cite{BHS} suggests a
five-loop divergence, unless additional cancellation mechanisms are
present.  

In the early 1980s, a seven-loop $\CN=8$ supersymmetric counterterm 
was constructed by Howe and Lindstr\"om~\cite{Howe1980th}.  
Also at that time, an eight-loop counterterm was
presented~\cite{Howe1980th,Kallosh1980fi}, which is manifestly invariant
under a continuous noncompact coset symmetry possessed by $\NeqEight$
supergravity, referred to as $E_{7(7)}$.  The status of lower-loop
counterterms with respect to $E_{7(7)}$ is not totally clear.
For example, the volume of the on-shell ${\cal N}=8$ superspace would 
appear at seven loops, and is invariant under $E_{7(7)}$, but it might also
vanish~\cite{BHS2}.

In the last few years, a variety of arguments have highlighted
the excellent ultraviolet properties of $\CN=8$ supergravity, mostly
suggesting finiteness until at least seven loops, although some arguments
go much further.  Based on multi-loop superstring results obtained by
Berkovits using the pure spinor formalism~\cite{Berkovits},
Green, Russo and Vanhove~\cite{GreenII} argued that the theory should be
finite through nine loops.  However, a recent re-analysis by the same
authors~\cite{GRV2010} (see also refs.~\cite{Vanhove2010,BG2010})
indicates that technical issues with the pure spinor formalism
beginning at five loops invalidate this argument, and suggest a first
divergence at seven loops.  Note that
refs.~\cite{Howe1980th,BHS2,KalloshLightCone}
also point to seven loops for a possible first divergence.

There are also arguments based on M-theory
dualities which suggest finiteness to all
orders~\cite{DualityArguments}.  On the other hand, $\CN=8$ supergravity
is a point-particle theory in four dimensions containing only massless
particles. String theory and M theory are quite different.
Perturbatively, they contain infinite towers of massive excited states
(string and/or Kaluza-Klein excitations); nonperturbatively, they contain
additional extended objects.  It is not at all obvious that results
found in those theories can be applied directly to $\CN=8$ supergravity,
because of issues in decoupling the infinite towers of states~\cite{GOS}.
The safest approach to determining the ultraviolet properties of
$\CN=8$ supergravity is to work directly in the field theory.

The purpose of these lectures is to describe some of the main methods
that have been used to determine multi-loop amplitudes in $\CN=8$
supergravity, and to extract from the amplitudes the ultraviolet
behavior of the theory. Another review which overlaps this one
in subject matter has been written recently 
by Bern, Carrasco and Johansson~\cite{BCJErice}.

Multi-loop computations in $\CN=8$ supergravity are feasible thanks to
two key ideas that work hand in hand with each other: 
the unitarity method~\cite{UnitarityMethod},
which allows loop computations to be reduced to tree computations;
and the Kawai-Lewellen-Tye (KLT) relations~\cite{KLT}, which allow $\CN=8$
supergravity tree amplitudes to be written in terms of tree amplitudes
in a gauge theory, $\CN=4$ super-Yang-Mills theory ($\CN=4$ SYM).
Both of these ideas have more modern refinements, to be described later,
which are useful for pushing the results to the maximal number of loops.
Using these methods, as I will explain in more detail below,
the four-graviton amplitude in $\CN=8$ supergravity
was computed at two loops in 1998~\cite{BDDPR}, and eventually
at three~\cite{GravityThree,CompactThree} and four loops~\cite{Neq84}.
The basic idea is to first compute the four-gluon
amplitude in $\CN=4$ SYM at the same number of loops, and then use
this information, along with unitarity and the KLT relations, to 
reconstruct the four-graviton amplitude in $\CN=8$ supergravity.

Much of the motivation for these multi-loop efforts came
from the results of one-loop computations with a large number of external
gravitons~\cite{OneloopMHVGravity,NoTriangle,NoTriangleB}.
The one-loop results all had the property (dubbed the ``no-triangle 
hypothesis''~\cite{NoTriangleB}) that the ultraviolet behavior
of $n$-graviton amplitudes in $\CN=8$ supergravity was
{\it never any worse} than that of the corresponding $n$-gluon amplitudes in
$\CN=4$ SYM.  Both sets of one-loop amplitudes are finite in four
dimensions.  Considered as theories with the same number of supercharges
in a higher space-time dimension, they first begin to diverge in
eight dimensions.
It is possible to use unitarity to argue that the one-loop behavior
also implies large classes of multi-loop cancellations~\cite{Finite}, 
although it clearly does not control the complete
multi-loop behavior.

Remarkably, the same property, that $\CN=8$ supergravity is no worse
behaved than $\CN=4$ SYM, has also been found to hold for
all the multi-loop four-graviton results that have been computed to
date, all the way through four loops.
Now $\CN=4$ SYM is well known to be an ultraviolet-finite theory
to all orders in perturbation theory~\cite{Neq4Finite};
indeed, it is the prototypical four-dimensional conformal field theory.
As long as $\CN=8$ supergravity is as well-behaved as $\CN=4$ SYM is
in the ultraviolet, it must also remain finite.  It is still an open
question as to whether this property holds to all orders in perturbation
theory.  It has been suggested~\cite{BG2010} that it might fail as soon
as the next uncomputed amplitude, namely at five loops.
(Such a failure at five loops could indicate~\cite{BG2010} a
four-dimensional divergence as early as seven loops, a loop
order also suggested in refs.~\cite{Howe1980th,BHS2,KalloshLightCone}.)
However, at
present it seems that only a complete computation of the five-loop
amplitude can definitively answer this question.

The remainder of this article is organized as follows.  In
\sect{DivQGSection} the general structure of divergences in quantum
gravity and supergravity is outlined.
In \sect{ToolsSection} we describe the KLT relations and the
unitarity method.  We discuss the method of maximal cuts, and the
rung rule for $\NeqFour$ super-Yang-Mills theory.
In \sect{KLTCopyingSection} we explain how the KLT relations
allow generalized cuts in $\NeqEight$ supergravity to be obtained
from two copies of the cuts in $\NeqFour$ super-Yang-Mills theory,
and we discuss the ultraviolet behavior of typical contributions.
\Sect{Neq8ThreeLoopSection} discusses the four-point scattering amplitude
and ultraviolet behavior of $\NeqEight$ supergravity at three loops,
and \sect{Neq8FourLoopSection} does the same at four loops.
In \sect{ConclusionsSection} conclusions are presented, along with
an outlook for the future.

%%%%%%%%%%%%%%%%%%%%%%%%%%%%%%%%%%%%%%%%%%%%%%

\section{Divergences in quantum gravity}
\label{DivQGSection}

In general, divergences in quantum field theory are
associated with local operators, or {\it counterterms}.  Divergences
in off-shell quantities, such as off-shell two-point functions, can depend
on the gauge.  However, the ultraviolet divergences in on-shell scattering
amplitudes, which we shall be concerned with here, are gauge
invariant, {\it i.e.}~diffeomorphism invariant.  Therefore they should
be constructed from tensors that transform homogeneously under coordinate
transformations.  For scattering amplitudes containing only external
gravitons, counterterms should be built from products of the Riemann
tensor $R_{\mu\nu\sigma\rho}$, including contractions of it such as the
Ricci tensor $R_{\mu\nu}$ and the Ricci scalar $R$.  

The available counterterms are also constrained by dimensional analysis.
The loop-counting parameter $G_N$ has mass dimension $-2$, so 
at each successive loop order a logarithmic divergence must be 
associated with a counterterm with a dimension two units greater
than in the previous loop. The Riemann tensor has mass dimension 2:
$R^\mu_{\nu\sigma\rho} \sim \del_\rho \Gamma^\mu_{\nu\sigma}
\sim g^{\mu\kappa} \del_\rho\del_\nu g_{\kappa\sigma}$.
Because the (tree-level) Einstein Lagrangian $R$ also has dimension 2,
an $L$-loop counterterm should have the schematic form 
${\cal D}^{2k} R_{\ldots}^{L+1-k}$, where ${\cal D}$ is a covariant
derivative and $R_{\ldots}$ is a generic Riemann tensor.

Even for on-shell divergences, there is still an ambiguity in associating
counterterms, because of the freedom to perform (nonlinear) field
redefinitions on the action.  For example, in the case of pure 
Einstein gravity, with action $S = \int d^4x \sqrt{-g} R$, one can redefine
the metric by $g_{\mu\nu} \to f(g) g_{\mu\nu}$.  Because the variation
$\delta S/\delta g_{\mu\nu}$ is proportional to the Ricci tensor
$R_{\mu\nu}$, one can adjust in this way the coefficient of any potential
counterterm in pure Einstein gravity that contains $R_{\mu\nu}$. 
At one-loop, the potential counterterm
$R_{\mu\nu\sigma\rho}R^{\mu\nu\sigma\rho}$ is thus equivalent to
$R^2 - 4 R_{\mu\nu}R^{\mu\nu} +R_{\mu\nu\sigma\rho}R^{\mu\nu\sigma\rho}$,
which is the Gauss-Bonnet term, a total derivative in four dimensions.
Total derivatives cannot be produced in perturbation theory.
For this reason, pure Einstein gravity is ultraviolet-finite at one
loop~\cite{tHooftVeltmanGravity}.
On the other hand, if matter is coupled to gravity, there are 
generically counterterms of the
form $T_{\mu\nu}T^{\mu\nu}$, where $T^{\mu\nu}$ is the matter stress-energy
tensor, and the theory is nonrenormalizable at one
loop~\cite{tHooftVeltmanGravity}.

At two loops, there is a nontrivial counterterm for pure gravity,
which cannot be related to a total derivative.  It has the form,
\be
R^3 \equiv 
R^{\lambda\rho}_{\mu\nu} R^{\mu\nu}_{\sigma\tau}
R^{\sigma\tau}_{\lambda\rho} \,.
\label{R3def}
\ee
Two independent two-loop computations using Feynman diagrams,
by Goroff and Sagnotti~\cite{GoroffSagnotti} and by van de
Ven~\cite{vandeVen}, showed that the coefficient of this counterterm is
nonzero, {\it i.e.}~that pure gravity diverges at two loops.

In any supergravity, even ${\cal N}=1$, if all states are in the
same multiplet with the graviton (pure supergravity), then it is possible
to show that there are no divergences until at least three
loops~\cite{Grisaru,DeserKayStelle,Tomboulis}.
This property can be understood by computing the tree-level matrix element
of the operator $R^3$ between four on-shell graviton states with
outgoing helicity $({-}{+}{+}{+})$~\cite{Grisaru}.  On the one hand, the
matrix element is nonzero~\cite{vanNWu}.  On the other hand,
such a helicity configuration is forbidden by supersymmetric Ward
identities acting on the $S$ matrix~\cite{GrisaruSWI}
(see \sect{RungRuleSubsection}), which
require at least two negative and two positive helicities.
Thus $R^3$ is unavailable as a potential two-loop counterterm.
In a pure supergravity, all four-point amplitudes can be rotated by
supersymmetry into four-graviton amplitudes, so they must also be
finite.

At three loops, the first potential counterterm in pure supergravity
arises, 
\be
R^4
\ \equiv\ t_8^{\m_1\n_1\ldots\m_4\n_4} t_8^{\r_1\sigma_1\ldots\r_4\sigma_4}
R_{\m_1\n_1\r_1\sigma_1} R_{\m_2\n_2\r_2\sigma_2}
R_{\m_3\n_3\r_3\sigma_3} R_{\m_4\n_4\r_4\sigma_4} \,,
\label{R4def}
\ee
where the tensor $t_8$ is defined in ref.~\cite{SchwarzReview}.
This particular operator~\cite{DeserKayStelle,Ferrara1977mv,Deser1978br,%
Howe1980th,Kallosh1980fi}
is also known as the square of the Bel-Robinson
tensor~\cite{Bel1958}.  It is compatible with supersymmetry, not just
${\cal N}=1$ but all the way up to maximal $\CN=8$ supersymmetry.
This latter property follows from the appearance of $R^4$ in the 
low-energy effective action of the $\CN=8$ supersymmetric
closed superstring~\cite{Gross1986iv}; indeed, it represents 
the first correction term beyond the limit of
$\CN=8$ supergravity~\cite{Green1982sw},
appearing at order $\alpha^{\prime3}$.  More precisely,
the matrix element of this operator on four-particle states
takes the form,
\be
\langle R^4 \rangle |_{4-{\rm point}} =
s t u \, M_4^{\tree}(1,2,3,4) \,,
\label{R4matrixelement}
\ee 
where the momentum invariants are
$s=(k_1+k_2)^2$, $t=(k_2+k_3)^2$, $u=(k_1+k_3)^2$, and
$M_4^\tree$ stands for any of the $256^4$ four-point amplitudes in
${\cal N}=8$ supergravity (stripped of the gravitational 
coupling constant).

As mentioned earlier, it was generally believed prior to 1998
that the $R^4$ counterterm would control the first divergence
in $\NeqEight$ supergravity.  However, when the two-loop four-graviton
amplitude was computed~\cite{BDDPR}, it was found to have two
additional powers of the momentum-invariants emerging from the integrand,
so that the amplitude had the schematic form,
\be
M_4^{\twoloop}(1,2,3,4) =
s t u \, M_4^{\tree}(1,2,3,4) \times 
\Bigl[ s^2\ I_4^X(s,t)\ +\ \hbox{permutations} \Bigr] \,,
\label{twoloopschematic}
\ee
where $I_4^X(s,t)$ denotes a scalar double box integral.
The two extra powers of $s$ in \eqn{twoloopschematic},
when written in position space,
correspond to a counterterm with four covariant derivatives,
of the form ${\cal D}^4R^4$, where we have not specified
the precise index contractions.  This operator has a dimension
appropriate for a four-dimensional divergence at five loops,
not three loops.  Ref.~\cite{BDDPR} also investigated a few
different classes of higher-loop contributions, and found that in each
case, two extra powers of momentum-invariants continued to emerge 
from the loops.  As it later transpired, the true power-counting for
the three- and four-loop amplitudes is considerably better, due to
cancellations between different contributions.  As we shall see,
the actual behavior observed so far is ${\cal D}^6R^4$ at three
loops~\cite{GravityThree,CompactThree,BG2010} and ${\cal D}^8R^4$
at four loops~\cite{Neq84,BG2010}.

For simplicity, I have represented the ultraviolet behavior at
a given loop order in terms of local counterterms,
even when the theory is finite in four dimensions.
This terminology corresponds to considering $\NeqEight$ supergravity
generalized to $D\geq4$ dimensions, while keeping the same number of
supercharges as in four dimensions (thirty-two).  The same 256
physical states circulate in the loop for any value of $D$,
but the virtual loop integration measure $d^D\ell$ leads
to poorer ultraviolet behavior as $D$ increases.
At each loop order $L$, there is a critical dimension
$D_c(L)$ at which the four-point amplitude first diverges in
the ultraviolet, and this can be related to a counterterm
of the form ${\cal D}^{2k}R^4$.  The gravitational coupling
$\kappa$ is related to Newton's constant by $\kappa^2 = 32 \pi^2 G_N$.
In $D$ dimensions, $\kappa^2$ (the loop-counting parameter) has
a mass dimension of $-(D-2)$.  A counterterm of the form 
${\cal D}^{2k}R^4$ corresponds to a logarithmic amplitude 
divergence of the form
\be
{\cal M}_4^{L{\rm \hbox{-}loop}}
\ \sim\ \kappa^{2L} \times stu {\cal M}_4^\tree \times s^k \times
\biggl[ \frac{1}{\e}\ \hbox{or}\ \ln\Lambda \biggr] \,,
\label{DdimDivergence}
\ee
when using dimensional regularization ($D \to D-2\e$) or a dimensionful
cutoff $\Lambda$.  Dimensional analysis on \eqn{DdimDivergence}
then implies that
\be
2k+6\ =\ L(D_c-2).
\label{kvsDc}
\ee
Suppose $k=2$ at each loop order beyond $L=1$, the behavior suggested in
ref.~\cite{BDDPR}.  (The one-loop case is special, as no extra powers
of $s$ emerge at this order~\cite{Green1982sw}, giving $D_c(1)=8$.)
This behavior would imply
\be
D_c(L)\ =\ 2 + \frac{10}{L}\qquad (L>1),
\label{OldPowerCount}
\ee
yielding a divergence in four dimensions at $L=5$.  On the other hand,
the behavior $k=L$ is seen through four loops, and this would imply,
if valid to all loops,
\be
D_c(L)\ =\ 4 + \frac{6}{L}\qquad (L>1).
\label{NewPowerCount}
\ee
Because this formula gives a critical dimension above 4 for all $L$, 
its validity would result in the perturbative finiteness of 
$\NeqEight$ supergravity.

It is convenient to consider simultaneously the ultraviolet
behavior of $\NeqFour$ SYM generalized to $D\geq4$.
Ref.~\cite{BDDPR} studied the four-gluon amplitude in this
theory~\cite{BRY}, the complete amplitude at two loops
(see also ref.~\cite{MarcusSagnotti}),
and classes of higher-loop contributions.  In this case,
a single additional power of $s$ emerged, beyond that at one loop,
leading to the generic expected behavior, 
\be
{\cal A}_4^{L{\rm \hbox{-}loop}}
\ \sim\ g^{2L} \times st {\cal A}_4^\tree \times s \times
\biggl[ \frac{1}{\e}\ \hbox{or}\ \ln\Lambda \biggr] \,.
\label{DdimDivergenceSYM}
\ee
In $D$ dimensions, $g^2$ (the gauge theory loop-counting parameter)
has a mass dimension of $-(D-4)$.  Dimensional analysis on
\eqn{DdimDivergenceSYM} then implies $4+2 = L(D_c-4)$, or 
$D_c(L) = 4+6/L$, the same behavior as in \eqn{NewPowerCount}.  The
associated higher-dimensional counterterms have the form ${\cal D}^2 F^4$,
where $F = F^{\mu\nu}$ is the gauge field strength and ${\cal D}$ a
gauge covariant derivative.

This multi-loop behavior of $\NeqFour$ SYM
follows from an argument by Howe and Stelle~\cite{HoweStelleNew}
using ${\cal N}=3$ harmonic superspace~\cite{HarmonicSuperspace}.
In the limit of a large number of colors, {\it i.e.}~for gauge group
$SU(N_c)$ for large $N_c$, the behavior has been confirmed by explicit
computation through five loops, using the planar amplitudes in
refs.~\cite{BDS,FourLoop,FiveLoop}.
In other words, at the four-point level, at each loop order,
\eqn{NewPowerCount} is equivalent to the statement
that the ultraviolet behavior of $\NeqEight$ supergravity is 
no worse than that of $\NeqFour$ super-Yang-Mills theory.

There have been suggestions~\cite{BG2010,Vanhove2010} recently
that the four-point five-loop amplitude in $\NeqEight$ supergravity
will behave as ${\cal D}^8R^4$, not ${\cal D}^{10}R^4$.
This behavior would be worse than that of $\NeqFour$ super-Yang-Mills
theory.  Although it would not directly imply a four-dimensional
divergence at five loops, it would have important implications for the
possibility of finiteness to all orders.  However, it seems that
only a complete computation of the amplitude can definitively
answer this question.

I have not been very specific about the precise index contractions
for the various counterterms, nor about the number of independent
counterterms at each dimension (or value of $k$ in ${\cal D}^{2k}R^4$).
Also, what about the possibility of counterterms that do not show up
at the four-point level at all?  Recently, Elvang, Freedman and Kiermaier
used $\NeqEight$ supersymmetry and locality to heavily constrain the
possibilities~\cite{EFK2010}.  At the four-point level, the constraints
on operators of the form ${\cal D}^{2k}R^4$ are that their matrix elements
should generalize \eqn{R4matrixelement} as follows,
\be
\langle {\cal D}^{2k}R^4 \rangle |_{4-{\rm point}} =
P_k(s,t,u) \, s t u \, M_4^{\tree}(1,2,3,4) \,,
\label{D2kR4matrixelement}
\ee 
where $P_k(s,t,u)$ must be a Bose symmetric polynomial of degree $k$,
subject to the on-shell constraint $s+t+u=0$.
It is easy to count the independent polynomials, and hence the number
of four-point counterterms allowed by $\NeqEight$ supersymmetry,
at loop order $L=k+3$.   There is one such operator for $k=0$;
none for $k=1$ (because $s+t+u=0$); one each for $k=2,3,4,5$; two
for $k=6$; and so on~\cite{EFK2010}.  More impressively,
ref.~\cite{EFK2010} also showed that {\it beyond} the four-point level
very few operators are allowed at low loop order; in fact, there are
no such operators until $L=7$.  (The four-loop case was analyzed
earlier~\cite{DHHK,Kallosh4loop}.)  Thus through six loops, all
ultraviolet divergences must appear in the four-graviton amplitude. 

%%%%%%%%%%%%%%%%%%%%%%%%%%%%%%%%%%%%%%%%%%%%%%

\section{Computational tools}
\label{ToolsSection}

In this section we outline the two major tools that have been used
to compute complete multi-loop amplitudes in $\NeqEight$ supergravity:
the KLT relations and the unitarity method.  We then discuss generalized
unitarity at the multi-loop level, the method of maximal cuts, and the
rung rule for $\NeqFour$ super-Yang-Mills theory.

\subsection{KLT relations}
\label{KLTSubsection}

%%%%%%%%%%%%%%%%%%%%%%% TABLE OF MULTIPLICITIES %%%%%%%%%%%%%%
\begin{table}[ht]
%\hspace{-2.5cm}
\begin{center}
\begin{tabular}{|c||c|c|c|c|c|c|c|c|c|}
\hline
\multicolumn{10}{|c|}{$\NeqEight$ supergravity} \\
\hline
$h$ & $-2$ & $-\textstyle{\frac{3}{2}}$ & $-1$ &
$-\textstyle{\frac{1}{2}}$ & $0$ & $\textstyle{\frac{1}{2}}$ 
& $1$ & $\textstyle{\frac{3}{2}}$ & $2$ \\  [2pt]
\hline
\# of states & $1$ & $8$ & $28$ & $56$ & $70$ & $56$ & $28$ & $8$ & $1$
\\
\hline
field & $h^-$ & $\psi_i^-$ & $v_{ij}^-$ & $\chi_{ijk}^-$
& $s_{ijkl}$ & $\chi_{ijk}^+$ & $v_{ij}^+$ & $\psi_i^+$ & $h^+$
\\ [2pt]
\hline
\hline
\multicolumn{10}{|c|}{$\NeqFour$ super-Yang-Mills} \\ 
\hline
$h$ & ~ & ~ & $-1$ &
$-\textstyle{\frac{1}{2}}$ & $0$ & $\textstyle{\frac{1}{2}}$ 
& $1$ & ~ & ~ \\ [2pt]
\hline
\# of states & ~ & ~ & $1$ & $4$ & $6$ & $4$ & $1$ & ~ & ~
\\
\hline
field &  &  & $g^-$ & $\lambda_A^-$ & $\phi_{AB}$
& $\lambda_A^+$ & $g^+$ & & 
\\ [2pt]
\hline
\end{tabular}
\end{center}
\caption{\small
Table of state multiplicities, as a function of helicity $h$,
for the $2^8=256$ states in $\NeqEight$ supergravity and for the 
$2^4 = 16$ states in $\NeqFour$ super-Yang-Mills theory. }
\label{MultiplicityTable} 
\end{table}

%%%%%%%%%%%%%%%%%%%%%%%%%%%%%%%%%%%%%%%%%%%%%%%%%%%%%%%%%%%

The $2^8 = 256$ massless states of $\NeqEight$ supergravity are tabulated
in \tab{MultiplicityTable}.  The multiplicity for helicity $h$ is
given by $({8\atop p})$, where $p=2(2-h)$ is the number of applications
of the spin-$1/2$ supersymmetry generators $Q_i$ needed to reach that
helicity from the $h=+2$ graviton state $h^+$.  Alternatively, the
multiplicities can be read off from the 8$^{\rm th}$ row of Pascal's
triangle, or the coefficients in the binomial expansion of $(x+y)^8$.
Also tabulated in \tab{MultiplicityTable} are the multiplicities for 
the $2^4 = 16$ massless states of $\NeqFour$ SYM.  They are given
by $({4\atop p})$ with $p=2(1-h)$, or as the coefficients in the
binomial expansion of $(x+y)^4$.
Both $\NeqEight$ supergravity and $\NeqFour$ SYM arise as the
low-energy limit of string theories~\cite{Green1982sw}, 
the type II closed superstring and the type I open superstring,
respectively.  The closed superstring contains
both left- and right-moving modes, which are each in correspondence with 
one copy of the the open superstring modes.  Therefore
the $\NeqEight$ supergravity Fock space can be written as the tensor 
product of two copies of the $\NeqFour$ SYM Fock space,
\be
[\NeqEight]\ =\ [\NeqFour]_L \, \otimes \, [\NeqFour]_R \,.
\label{Neq8from4}
\ee
The fact that the multiplicities work out is a simple consequence of
the identity $(x+y)^8 = [(x+y)^4]^2$.

%%%%%%%%% FIGURE %%%%%%%%%%%%%%%
\begin{figure}[t]
\centerline{\epsfxsize 4.5 truein \epsfbox{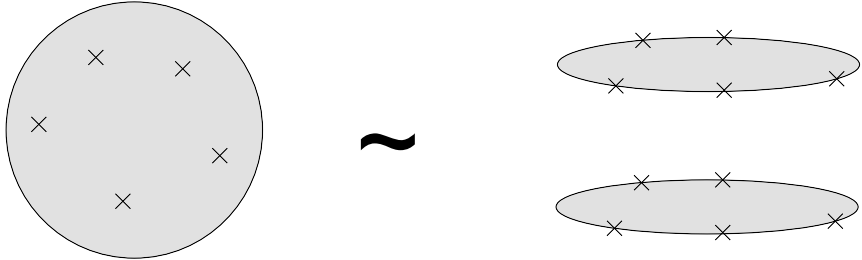}}
\caption[a]{\small Schematic depiction of the KLT relations.  The
closed-string world-sheet, a sphere, can be thought of as two copies
of the open-string world-sheet, a disk.  Vertex operator insertions are
marked with $\times$'s.
}
\label{KLTFigure}
\end{figure}
%%%%%%%%%%%%%%%%%%%%

Kawai, Lewellen and Tye~\cite{KLT} first observed that closed and open string
amplitudes are very closely related at tree level.  As shown in
\fig{KLTFigure}, the closed-string world-sheet is a sphere, and the
emission of a particular state is described by inserting a closed-string
vertex operator $V^{\rm closed}(z,\bar{z})$ somewhere on the sphere.
In contrast, the open-string world-sheet is a disk, and open string states
are emitted off the boundary, with a vertex operator $V^{\rm open}(x)$.
The KLT relations derive from the fact that the closed-string vertex
operator is a product of two open-string vertex operators,
\be
V^{\rm closed}(z_i,\bar{z}_i) = V_L^{\rm open}(z_i)\, 
\, \times \, \overline{V}_R^{\rm  open}(\bar{z}_i) \,,
\label{ClosedVertex}
\ee
one for the left-movers and one for the right-movers.
The left and right string oscillators appearing in $V_L^{\rm open}$ and
$\overline{V}_L^{\rm open}$ are distinct, but the zero-mode momentum
is shared.  Closed-string tree amplitudes are given by integrating
correlation functions of the form
\be
\langle V_1^{\rm closed}(z_1,\bar{z}_1)
 \cdots V_n^{\rm closed}(z_n,\bar{z}_n) \rangle
\label{ClosedTree}
\ee
over $n-3$ copies of the sphere.  KLT noticed that the
integrand~(\ref{ClosedTree}) was just the product of corresponding
open-string integrands.  They expressed the closed-string complex
integrations as products of $z$ and $\bar{z}$ contour integrals, 
and then deformed the contours until they were equivalent
to open-string integrals, over real variables $x_i$,
multiplied by momentum-dependent phase factors arising from
branch cuts in the integrand.
In this way, arbitrary closed-string tree amplitudes were
expressed as quadratic combinations of open-string tree amplitudes.

Taking the low-energy limit of the KLT relations for string theory gives
corresponding relations at the field-theory level, relating 
$n$-point tree amplitudes $M_n^\tree$ in $\NeqEight$ supergravity
to quadratic combinations of tree amplitudes $A_n^\tree$ in
$\NeqFour$ super-Yang-Mills theory.  In the low-energy limit,
the momentum-dependent phase factors generate powers of
momentum-invariants.  Strictly speaking, the gauge
theory tree amplitudes that appear are those from which the Chan-Paton
factors have been removed (in string terminology), or color-ordered
subamplitudes (in QCD terminology).  (For a review, see 
ref.~\cite{LDTASI}.)  We write the full tree amplitude as
\be
{\cal A}_n^\tree(\{k_i,a_i\}) = g^{n-2} \,
 \sum_{\rho\in S_n/\ZZ_n}
\Tr( T^{a_{\rho(1)}} T^{a_{\rho(2)}} \ldots T^{a_{\rho(n)}} ) \,
 A_n^\tree(\rho(1), \rho(2), \ldots, \rho(n))\,,
\label{Antreedef}
\ee
where $g$ is the gauge coupling, $a_i$ is an adjoint index,
$T^{a_i}$ is a generator matrix in the fundamental representation of
$SU(N_c)$, the sum is over all $(n-1)!$ inequivalent (non-cyclic)
permutations $\rho$ of $n$ objects, and the argument $i$
of $A_n^\tree$ labels both the momentum $k_i$ and state information
(helicity $h_i$, {\it etc.}).  In the case of supergravity amplitudes,
we only strip off powers of the coupling $\kappa$, defining $M_n^\tree$ by
\be
{\cal M}_n^\tree(\{k_i\})
\ =\ \left(\frac{\kappa}{2}\right)^{n-2}\,M_n^\tree(1,2,\ldots,n) \,.
\label{Mntreedef}
\ee

Then the first few KLT relations have the form,
\bea
M_3^\tree(1,2,3) &=& i \, A_3^\tree(1,2,3) \tilde{A}_3^\tree(1,2,3) \,,
\label{KLTThree} \\
M_4^\tree(1,2,3,4) &=&  
     - i s_{12} \, A_4^\tree(1,2,3,4) \, \tilde{A}_4^\tree(1,2,4,3)\,, 
\label{KLTFour} \\
M_5^\tree(1,2,3,4,5) &=& i s_{12} s_{34} \, A_5^\tree(1,2,3,4,5)
                                     \tilde{A}_5^\tree(2,1,4,3,5)
\ +\ \CP(2,3) \,,
\label{KLTFive} \\
M_6^\tree(1,2,3,4,5,6) &=& - i s_{12}s_{45} 
\, A_6^\tree(1,2,3,4,5,6) \nonumber\\
&&\quad
\times \left[ s_{35} \, \tilde{A}_6^\tree(2,1,5,3,4,6)
            + (s_{34}+s_{35}) \, \tilde{A}_6^\tree(2,1,5,4,3,6) \right]
\nonumber\\
&&\quad\  +\ \CP(2,3,4) \,,
\label{KLTSix}
\eea
where $s_{ij} \equiv (k_i+k_j)^2$, and 
``$+\,\CP$'' indicates a sum over the $m!$ permutations of the 
$m$ arguments of $\CP$.  Here $A_n^\tree$ indicates a tree amplitude
for which the external states are drawn from the left-moving Fock space
$[\NeqFour]_L$ in the tensor product~(\ref{Neq8from4}), while
$\tilde{A}_n^\tree$ denotes an amplitude from the right-moving copy
$[\NeqFour]_R$.

The KLT relations are quite general, in the sense that left- and
right-movers apparently do not need to be drawn from the particular
gauge theory $\NeqFour$ SYM
(or any truncation of it)~\cite{OtherKLT,BDHK10}. Furthermore,
a general `double-copy' formula for gravity amplitudes was
proposed recently~\cite{BCJ08,BCJ10,BDHK10}, which is consistent
with the KLT relations, but generates additional novel representations
of gravity amplitudes.  It starts with a representation~\cite{BCJ08}
of the full color-dressed gauge tree amplitudes in terms of cubic 
graphs labelled by $i$,
\be
\frac{{\cal A}_n^\tree}{g^{n-2}} 
\ =\ \sum_i \frac{n_i c_i}{ (\prod_j p_j^2)_i } \,,
\qquad\quad
\frac{{\cal \tilde{A}}_n^\tree}{g^{n-2}}
\ =\ \sum_i \frac{\tilde{n}_i c_i}{ (\prod_j p_j^2)_i } \,,
\label{cubicgaugetrees}
\ee
where $1/p_j^2$ are scalar propagators, and $n_i,\tilde{n}_i$
are kinematical numerator factors for the left- and right-moving theory. 
The $c_i$ are color factors, taken for convenience to be in a theory with
adjoint particles only, so that they are specific products of the
structure constants $f^{abc}$, one for each cubic vertex, and contracted
according to the topology of the graph.  The Jacobi identity,
\be
f^{abe} f^{cde} + f^{cae} f^{bde} + f^{bce} f^{ade} = 0,
\label{ActualJacobi}
\ee
induces relations between the color factors for $n$-point amplitudes,
namely
\be
c_i + c_j + c_k = 0,
\label{colorJacobi}
\ee
where the three graphs $i,j,k$ are identical except for exchanging
the connections between two of the vertices, following the
relation~(\ref{ActualJacobi}). 

While there is considerable freedom in choosing the numerator
factors, it seems that it is always possible to choose them to 
satisfy kinematic analogs of the color Jacobi identity~\cite{BCJ08},
\be
n_i + n_j + n_k = 0,
\label{kinematicJacobi}
\ee
for all triplets $(i,j,k)$ for which the color Jacobi
identity~(\ref{colorJacobi}) holds.
With such a choice made, the double-copy formula for the
gravity amplitudes is then
\be
M_n^\tree = i \sum_i \frac{ n_i \tilde{n}_i }{ (\prod_j p_j^2)_i } \,.
\label{cubicgravitytrees}
\ee
Although much of the multi-loop progress in $\NeqEight$ supergravity 
to date has been based on the KLT relations, it is quite likely that
representations such as \eqn{cubicgravitytrees} will play a key role in
the future.

These relations between gravity and gauge theory are reminiscent of the
AdS/CFT duality~\cite{AdSCFT}.  Of course, the details are very different:
AdS/CFT relates a weakly-coupled gravitational theory to a
strongly-coupled gauge theory, whereas KLT and associated relations related
a weakly-coupled gravitational theory to the {\it square} of a
weakly-coupled gauge theory.  AdS/CFT is tied to the notion of
holography.  Similarly, there is undoubtedly a deep principle attached
to KLT-like dualities, but its full nature has not yet been unraveled.

%%%%%%%%%%%%%%%%%%%%%%%%%%%%%%%%%%%%%%%%%%%%%%%%%%%%

\subsection{Unitarity method}
\label{UnitaritySubsection}

\begin{figure}[t]
\centerline{\epsfxsize 5.2 truein \epsfbox{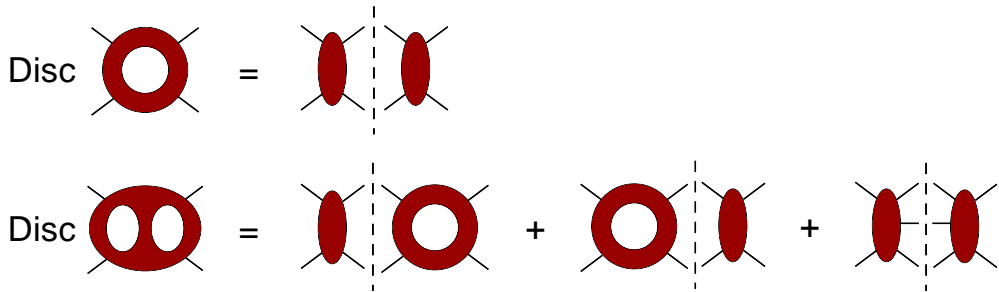}}
\caption[a]{\small Unitarity relations for the four-point amplitude
at one and two loops.  The number of holes in a blob indicates the number
of loops in the corresponding amplitude.}
\label{UnitarityFigure}
\end{figure}

The scattering matrix is a unitary operator between in and out states.
That is, $S^\dagger S = 1$, or in terms of the more standard
``off-forward'' part of the $S$ matrix, $T \equiv (S-1)/i$, 
we have
\be
2 \,\hbox{Disc} \, T = T^\dagger T \,,
\label{Tunit}
\ee
where Disc\,$T \equiv (T-T^\dagger)/2i$.
This simple relation gives rise to the well-known unitarity relations,
or cutting rules~\cite{Cutting},
for the discontinuities (or absorptive parts) of
perturbative amplitudes.  If one inserts a perturbative expansion for
$T$ into \eqn{Tunit}, say
\bea
T_4 &=& g^2 \, T_4^\tree + g^4 \, T_4^\oneloop + g^6 \, T_4^\twoloop +
\ldots \,,
\label{T4} \\
T_5 &=& g^3 \, T_4^\tree + g^5 \, T_5^\oneloop + g^7 \, T_5^\twoloop
+ \ldots \,,
\label{T5}
\eea
for the four- and five-point amplitudes, then one obtains the unitarity
relations shown in \fig{UnitarityFigure}.

%%%%%%%%% FIGURE %%%%%%%%%%%%%%%
%%%%%%%%%%%%%%%%%%%%

At order $g^4$, the discontunity
in the one-loop four-point amplitude is given by the product of two 
order $g^2$ four-point tree amplitudes.  The product must be summed over 
all possible intermediate states crossing the cut (indicated by 
the dashed line), and integrated over all possible intermediate momenta.
At two loops, or order $g^6$, there are two possible types of cuts:
the product of a tree-level and a one-loop four-point amplitude
($g^2 \times g^4$), and the product of two tree-level five-point
amplitudes ($g^3 \times g^3$). 

To get the complete scattering amplitude, not just the absorptive part,
one might attempt to reconstruct the real part via a dispersion relation.
However, in the context of perturbation theory, an easier method is
available, because one knows that the amplitude could have been calculated
in terms of Feynman diagrams.  Therefore it can be expressed as a linear
combination of appropriate Feynman integrals, with coefficients that are
rational functions of the kinematics.  The unitarity
method~\cite{UnitarityMethod} matches the information coming from the
cuts against the set of available loop integrals in order to determine
these rational coefficients.  There are also additive rational terms
in the amplitude, terms which have no cuts in four dimensions.
There are two general ways to determine
these terms:  one can use unitarity in $D-4-2\e$
dimensions~\cite{DDimUnitarity,MultiLoopDDimGenUnitarity},
or one can exploit factorization information to relate the rational
terms for an $n$-point amplitude to those for amplitudes with $(n-1)$
or fewer legs~\cite{ee4partons,BCFRecursion,Bootstrap}.

At one loop, there have been many recent refinements to these on-shell
methods, allowing for automation and numerical implementation by several
groups~\cite{OPP,EGKMZ,BlackHat} (as reviewed recently in
ref.~\cite{BergerFordeReview}).   These results have led in turn
to state-of-the-art results for next-to-leading order QCD cross
sections, providing precise predictions for important Standard Model
backgrounds at the LHC.

\begin{figure}[t]
\centerline{\epsfxsize 5.2 truein \epsfbox{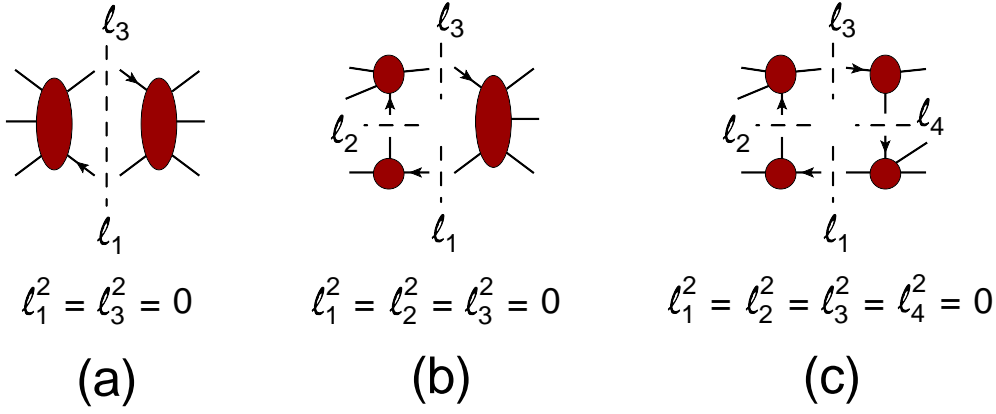}}
\caption[a]{\small Generalized unitarity at one loop: (a) the ordinary
  two-particle cut imposes two constraints on the loop-momentum; (b) the
  triple cut imposes three constraints and is sensitive to triangle
  coefficients; (c) the quadruple cut imposes four constraints (freezing all
  four components of $\ell_1^\mu$) and is sensitive to box coefficients.}
\label{GenU1lFigure}
\end{figure}

In particular, {\it generalized unitarity}~\cite{GeneralizedUnitarityOld},
which corresponds at one loop to cutting more than two lines, can be used 
to simplify the information required for computing many
terms in the amplitude~\cite{ee4partons,MoreGenUnitarity,BCFUnitarity}.
\Fig{GenU1lFigure}(a) depicts an ordinary cut for a one-loop six-point
amplitude in four space-time dimensions. 
The two lines crossing the cut, with momenta $\ell_1^\mu$ and
$\ell_3^\mu$, are on shell, so that $\ell_1^2 = \ell_3^2 = 0$.  Of course
$\ell_1^\mu$ and $\ell_3^\mu$ are not independent (they differ by the
fixed external momenta), so these
conditions represent two equations for the four components of $\ell_1^\mu$.
One can impose up to two more equations, leading to the more restrictive
cut kinematics in \fig{GenU1lFigure}(b) (the triple cut), and finally
\fig{GenU1lFigure}(c) (the quadruple cut).  For the latter condition,
$\ell_1^2 = \ell_2^2 = \ell_3^2 = \ell_4^2 = 0$, the number of equations
equals the number of unknowns, and so there are generically two discrete
solutions. The coefficient $d_j$ of a scalar box integral with the
indicated topology can be found~\cite{BCFUnitarity} simply by summing
over the product of the four tree amplitudes, evaluated for the two 
solutions labeled by $\sigma$,
\bea
 d_j &=& {1\over2} \sum_{\sigma=\pm}
 A_{n_1}^\tree(\ldots,-\ell^\sigma_1,\ell^\sigma_2)\,
 A_{n_2}^\tree(\ldots,-\ell^\sigma_2,\ell^\sigma_3)
\nonumber\\
&&\hskip0.8cm \times \,
 A_{n_3}^\tree(\ldots,-\ell^\sigma_3,\ell^\sigma_4)\,
 A_{n_4}^\tree(\ldots,-\ell^\sigma_4,\ell^\sigma_1)\,,
\label{boxcoeff}
\eea
where the ellipses stand for the external momenta.

%%%%%%%%% FIGURE %%%%%%%%%%%%%%%
%%%%%%%%%%%%%%%%%%%%

From \fig{GenU1lFigure} it is clear that as one imposes more
constraints, the required tree amplitudes have fewer legs, so they will
become simpler.  On the other hand, the kinematic conditions
for generalized cuts are more constraining, and often cannot be satisfied
by real Minkowski momenta.  The limiting case of this is the three-point
amplitude. When all three legs are massless, the only solution to
\be
k_1^2 = k_2^2 = k_3^2 = 0, 
\qquad k_1^\mu + k_2^\mu + k_3^\mu = 0,
\label{threeptkin}
\ee
with real momenta $k_i^\mu$, is for all three momenta to be parallel,
such that $k_1^\mu = -\alpha k_3^\mu$, $k_2^\mu = -(1-\alpha) k_3^\mu$, 
for some real number $\alpha$.  This configuration is pathological
because all kinematic invariants vanish.  Clearly, the momentum-invariants
$s_{ij}$ all vanish: $s_{12}=(k_1+k_2)^2 = k_3^2 = 0$,
and similarly $s_{23}=s_{13}=0$.

One can also construct kinematic invariants from Weyl spinors 
based on the particle momenta.  Defining 
$\lambda_i^\alpha \equiv |k_i^+\rangle \equiv {1\over2}(1+\gamma_5)u(k_i)$
and 
$\tilde\lambda_i^{\dot\alpha} \equiv |k_i^-\rangle
\equiv {1\over2}(1-\gamma_5)u(k_i)$,
the spinor products are
\bea
\spa{i}.{j} &=& \pol_{\alpha\beta} \lambda_i^\alpha \lambda_j^\beta
= \langle k_i^-|k_j^+\rangle \,, \label{spadef} \\
\spb{i}.{j} &=& \pol_{\dot\alpha\dot\beta}
 \tilde\lambda_i^{\dot\alpha} \tilde\lambda_j^{\dot\beta}
= \langle k_i^+|k_j^-\rangle \,. \label{spbdef}
\eea
They satisfy
\be
\spb{i}.{j}\spa{j}.{i}
= \Tr[\textstyle{\frac{1}{2}}(1+\gamma_5)\ksl_i \ksl_j] = s_{ij} 
\,.
\label{absid}
\ee
For real momenta, $\spa{i}.{j}$ and $\spb{i}.{j}$ are complex conjugates
of each other.  Therefore they are complex square roots of $s_{ij}$,
\be
\spa{i}.{j} = e^{i\phi_{ij}} \sqrt{s_{ij}}\,, \qquad
\spb{i}.{j} = -e^{-i\phi_{ij}} \sqrt{s_{ij}}\,, 
\label{spabsqrts}
\ee
for some phase angle $\phi_{ij}$, which means that they too vanish
for three-point kinematics.

However, for complex momenta $k_i$ there is another type of solution:
If we choose all three negative-helicity two-component spinors to be
proportional,
\be
\tilde\lambda_1^{\dot\alpha}\ \propto\ \tilde\lambda_2^{\dot\alpha}
\ \propto\ \tilde\lambda_3 ^{\dot\alpha} \,,
\label{spbvanish}
\ee
then according to \eqn{spbdef} we have $\spb1.2=\spb2.3=\spb1.3=0$,
but the other three spinor products, $\spa1.2$, $\spb2.3$, and $\spb1.3$,
are allowed to be
nonzero (consistent with \eqn{absid}).  Hence an amplitude built
solely from $\spa{i}.{j}$ is nonzero and finite for this choice of 
kinematics. The three-gluon amplitude with two negative and one
positive helicity is such an object,
\be
A_3^\tree(1^-,2^-,3^+) = i \, { {\spa1.2}^4\over\spa1.2\spa2.3\spa3.1 } \,,
\label{mmp}
\ee
where we assign helicities with an all-outgoing convention ({\it i.e.}~if
a particle is incoming, it has a physical helicity opposite from its
helicity label).  In contrast, the parity conjugate of this amplitude,
\be
A_3^\tree(1^+,2^+,3^-) = -i \, { {\spb1.2}^4\over\spb1.2\spb2.3\spb3.1 } \,,
\label{ppm}
\ee
vanishes for the kinematics~(\ref{spbvanish}), but is nonzero and finite
for the conjugate kinematics satisfying,
\be
\lambda_1^{\alpha}\ \propto\ \lambda_2^{\alpha}
\ \propto\ \lambda_3 ^{\alpha} \,.
\label{spavanish}
\ee
%

%%%%%%%%%%%%%%%%%%%%%%%%%%%%%%

\subsection{Multi-loop generalized unitarity and maximal cuts}
\label{MultiLoopGenUSubsection}

\begin{figure}[t]
\centerline{\epsfxsize 5.2 truein \epsfbox{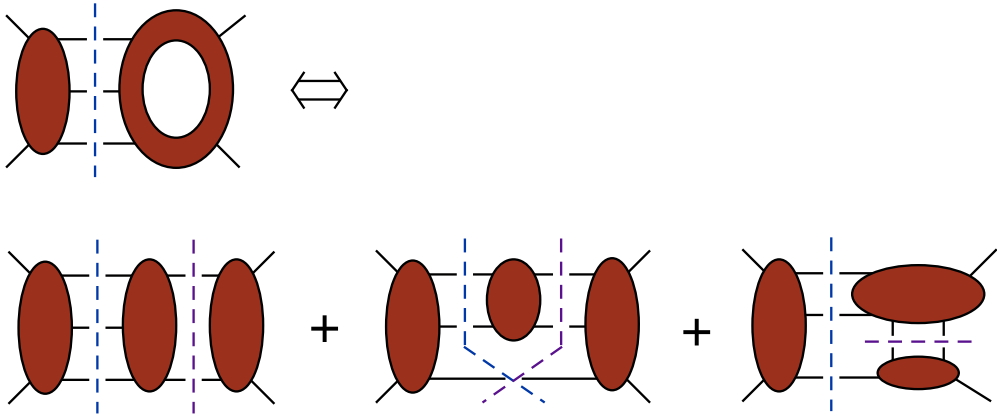}}
\caption[a]{\small An example of multi-loop generalized unitarity.
The one-loop five-point amplitude, appearing on the right side of
the ordinary cut, is further cut into products of trees,
in three inequivalent ways.}
\label{GenUMLFigure}
\end{figure}

At the multi-loop level, generalized unitarity is also
extremely powerful~\cite{MultiLoopDDimGenUnitarity,FourLoop,%
GravityThree,FiveLoop,LeadingSingularity}.
It allows one to avoid cuts like the ones shown in \fig{UnitarityFigure},
in which a loop amplitude appears on one side of the cut.
One can impose additional cuts in order to chop such loops further into
trees.  Doing so is essential in order to make use of the KLT
relations, which hold only for tree amplitudes.  
\Fig{GenUMLFigure} illustrates this, starting with an ordinary
three-particle cut for the three-loop four-point amplitude.
The information in this cut can be extracted more easily by cutting
the one-loop five-point amplitude on the right-hand side of the cut, 
decomposing it into the product of a four-point tree and a 
five-point tree; as illustrated, there are three inequivalent
ways to do this. 

%%%%%%%%% FIGURE %%%%%%%%%%%%%%%
%%%%%%%%%%%%%%%%%%%%

If one finds a representation of the amplitude 
that reproduces all the generalized cuts (in $D$ dimensions), then that 
representation is guaranteed to be correct.  The reason is that the
generalized cuts simply provide a way of efficiently sorting all the 
Feynman diagrams contributing to an amplitude, and every Feynman diagram
has a cut in some channel.  This statement assumes that all particles
are massless --- Feynman diagrams for external wave-function corrections
do not contain cuts, but they also vanish in dimensional regularization
in the massless case, because there is no scale on which they can depend.
The reason $D$ dimensions is
required is that some cuts may vanish as $D \to 4$.  An alternate but
equivalent argument uses dimensional analysis:  An $L$-loop 
amplitude in $D=4-2\e$ carries a fractional mass dimension of $-2\e L$,
from the integration measure $\sim (\int d^{4-2\e}\ell)^L$.
If there are no particle masses, then all dimensions are carried by
momentum invariants, corresponding to channels that can be cut through. 
A rational function $R(s,t)$ that is present in the amplitude in the
limit $\e\to0$ must actually have the form in $D=4-2\e$ of
$R(s,t)\times (-s)^{-\e L} \approx R(s,t)\times ( 1 - \e L \ln(-s))$.
The logarithm indicates that the rational function is visible in the cuts
at the next order in $\e$.

\begin{figure}[t]
\centerline{\epsfxsize 5.2 truein \epsfbox{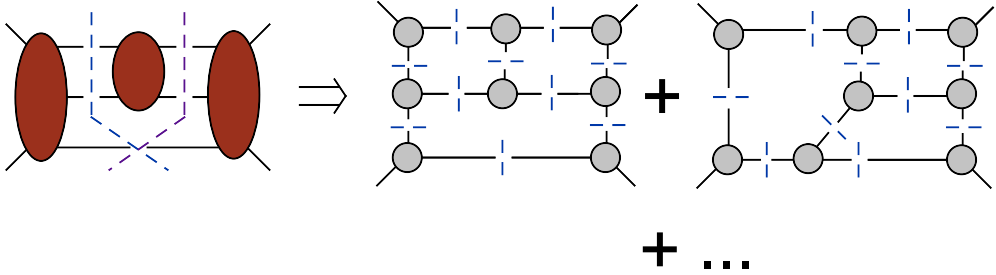}}
\caption[a]{\small Example of a real-momentum generalized cut generating
several maximal cuts; the latter contain only three-point tree amplitudes.}
\label{Cut3ToMaximalFigure}
\end{figure}

\Fig{GenUMLFigure} illustrates a particular type of generalized
unitarity in which all cut momenta are real.  As at one loop, it is
profitable to allow for complex momenta, and slice the tree amplitudes 
into yet smaller ones.  The method of
{\it maximal cuts}~\cite{FiveLoop,CompactThree} starts with
the limiting case in which all tree amplitudes are three-point ones.
\Fig{Cut3ToMaximalFigure} shows how one of the real-momentum configurations in
\fig{GenUMLFigure} spawns several maximal cuts.  The maximal cuts are
simply enumerated by drawing all cubic graphs.  Their evaluation is
also simple, because the three-point tree amplitudes are so compact.
In the case of gauge theory, they are given by \eqns{mmp}{ppm}, and by
other formulae related by supersymmetry; for gravity, they are obtained
by squaring the gauge amplitudes, using \eqn{KLTThree}.

%%%%%%%%% FIGURE %%%%%%%%%%%%%%%
%%%%%%%%%%%%%%%%%%%%

Even though the maximal cuts are maximally simple, they give an excellent
starting point for constructing the full amplitude.  For example,
for the four-gluon amplitude in $\NeqFour$ SYM, they detect all terms 
in the complete answer through two loops~\cite{BRY,BDDPR}, 
and all terms in the planar (leading in $N_c$) contribution through
three loops~\cite{BDS}.  The terms they do not detect can be
expressed as ``contact terms''.  Suppose we take the Feynman integral
associated with a cubic graph from scalar $\phi^3$ theory, and 
insert into the numerator one power of an inverse propagator
$\ell_i^2$ for some loop momentum $\ell_i$.  This insertion
cancels a propagator in the $\phi^3$ graph, which corresponds
to deleting one of the cut lines in the corresponding maximal cut,
and merging two of the three-point amplitudes in that cut
into a four-point amplitude.  By definition, such a contribution is
not detectable in the maximal cut, which assumes $\ell_i^2=0$.
However, it, and indeed all remaining terms in the amplitude,
can be found systematically, by considering the {\it near-maximal cuts}, 
which are found by collapsing one or more propagators in each 
maximal cut.  Starting with an ansatz based on the maximal cuts, one
adds contact terms to it, fixing their coefficients by requiring that
the new ansatz reproduces the near-maximal cuts.  The procedure is then
iterated, in the number of cancelled propagators, until all cuts are
successfully reproduced.  In the case of $\NeqFour$ SYM at $L$ loops,
the procedure converges after only $(L-2)$ cancelled propagators,
because of the excellent ultraviolet behavior of this theory,
\eqn{NewPowerCount}, which corresponds to only allowing $2(L-2)$
powers of loop momentum in the numerator of each cubic loop integral.

%%%%%%%%%%%%%%%%%%%%%%%%%%%%%%

\subsection{Supersymmetric Ward identities and the rung rule}
\label{RungRuleSubsection}

Although the maximal cuts are quite simple, there is an even simpler
subclass of contributions for maximally supersymmetric theories, 
those which contain iterated two-particle cuts,
{\it i.e.}~graphs that can be reduced to tree amplitudes by a succession
of two-particle cuts.  The reason they are so simple is two-fold:
(1) at each stage one encounters only four-point amplitudes, whose
dependence on the external states is completely dictated by supersymmetry,
at any loop order; and (2) the sum over intermediate states for this case
can be performed simply, once and for all.

The first statement follows from supersymmetric Ward
identities~(SWI) for the $S$ matrix~\cite{GrisaruSWI}.
These identities are derived by requiring that
supercharges annihilate the vacuum.  The ${\cal N}=1$ SWI
can be found easily by letting $Q = Q^\alpha\eta_\alpha$, 
where $\eta$ is a Grassmann parameter, and writing
\be
0 = \langle0|[Q,\Phi_1\Phi_2\cdots\Phi_n]|0\rangle
  = \sum_{i=1}^n \langle0|\Phi_1\cdots[Q,\Phi_i]\cdots\Phi_n]|0\rangle
\,,
\label{BasicSWI}
\ee
where $\Phi_i$ are fields making the $n$ external states for the
amplitude.  The commutators $[Q,\Phi_i] \equiv \tilde\Phi_i$ make
the corresponding ${\cal N}=1$ superpartner states.
If the $\Phi_i$ are chosen to make helicity eigenstates, and
in particular if many of the states are gluons with the same helicity,
then many terms in the SWI vanish~\cite{GrisaruSWI,LDTASI}.  

For an amplitude containing only gluons, in which all, or all but one, 
of the gluons have positive helicity, one can arrange that there
is only one term in the SWI, so that the amplitude itself vanishes,
\be
A_n^{\rm SUSY}(1^\pm,2^+,3^+,\ldots,n^+) = 0 \,.
\label{vanishSWI}
\ee
(The same argument applied to gravity implies the vanishing of the
four-point amplitude $M_4^{\rm SUSY}({-}{+}{+}{+})$; this fact was
used the argument in \sect{DivQGSection} for the absence of the $R^3$
counterterm in pure supergravity.)
A second type of vanishing amplitude contains $(n-2)$ gluons, plus
a single pair of states $(\overline{P},P)$:
\be
A_n^{\rm SUSY}(1_{\overline{P}}^{-h_P},2_P^{h_P},3^+,\ldots,n^+) = 0 \,.
\label{vanishmatterSWI}
\ee
Here $P$ may be a scalar $\phi$, a gluino $\lambda$, or a gluon 
$g$ (reverting to the previous case), with helicity
$h_P=0,\pm1/2,\pm1$ respectively:

The first nonvanishing class of $n$-point amplitudes are called
maximally-helicity-violating (MHV).  They include the pure-gluon amplitudes
with precisely two negative helicities (labeled by $i$ and $j$),
which were first written down at tree level by Parke and
Taylor~\cite{ParkeTaylor},
\be
A_n^\tree(1^+,\ldots,i^-,\ldots,j^-,\ldots,n^+) =
i \, { {\spa{i}.{j}}^4 \over \spa1.2\spa2.3\cdots\spa{n}.1 } \,.
\label{MHVPT}
\ee
Other MHV amplitudes include those with a pair of states 
$(\overline{P},P)$ as above, plus exactly one negative-helicity 
gluon (labelled by $j$). The SWI relate all such amplitudes to each
other, according to
\be
A_n^{\rm SUSY}(1_{\overline{P}}^{-h_P},2_P^{h_P},3^+,\ldots,j^-,\ldots,n^+)
 = \biggl( { \spa1.{j} \over \spa2.{j} } \biggr)^{2 h_P}
A_n^\tree(1_{\bar\phi},2_\phi,3^+,\ldots,j^-,\ldots,n^+) \,.
\label{MHVSWI}
\ee

Comparing the two cases in which $P$ is a gluon, $h_P=\pm1$, and repeating
for other labelings, \eqn{MHVSWI} implies that
the dependence of the MHV pure-gluon amplitudes on the location of the 
negative helicities is trivial,
\be
A_n^{\NeqFour\ \rm SYM}(1^+,\ldots,i^-,\ldots,j^-,\ldots,n^+) =
{\spa{i}.{j}}^4 \times f_n(s_{m,m+1}) \,,
\label{MHVpureg}
\ee
where $f_n(s_{m,m+1})$ is invariant under cyclic permutations,
and is completely independent of $i$ and $j$.
\Eqn{MHVpureg} is clearly satisfied by the Parke-Taylor tree
amplitudes~(\ref{MHVPT}), but it holds to arbitrary loop order in
$\NeqFour$ SYM.  At the four-point level, all nonvanishing amplitudes
are MHV, and so they are all related by $\NeqFour$ supersymmetry.
The same factor ${\spa{i}.{j}}^4$ occurs at every loop order,
and hence the ratio $A_n^{(L)}/A_n^\tree$ is independent of
$i$ and $j$ in $\NeqFour$ SYM.

\begin{figure}[t]
\centerline{\epsfxsize 4.5 truein \epsfbox{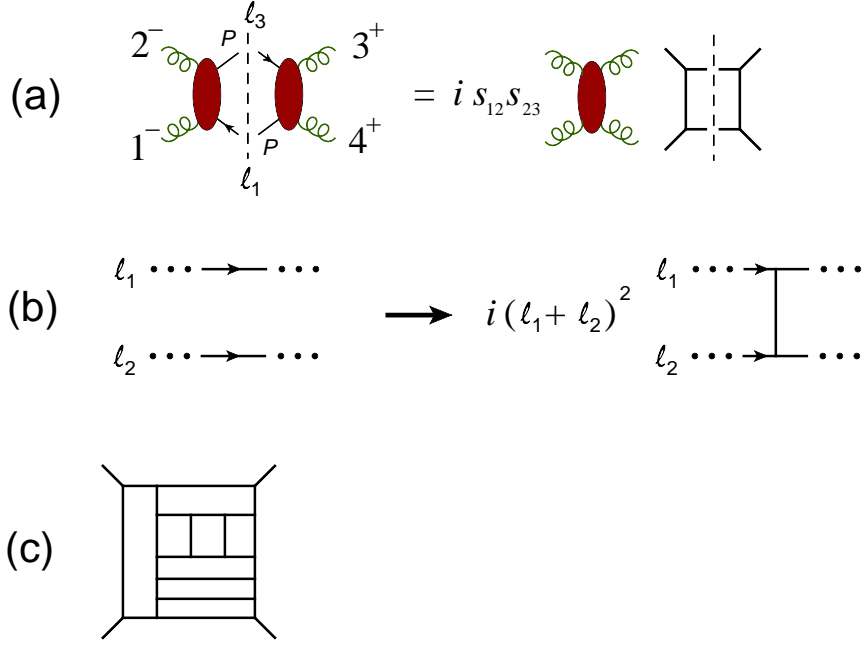}}
\caption[a]{\small (a) The $s$-channel cut of the one-loop four-gluon
amplitude in $\NeqFour$ SYM is expressed in terms of the tree amplitude,
the cut scalar box integral, and external momentum invariants.
(b) The rung rule for generating numerator factors using iterated
two-particle cuts.  (c) Example of a graph whose numerator factor
can be determined by the rung rule.}
\label{RR1lFigure}
\end{figure}

With this information in hand, we consider the simplest unitarity cut
in $\NeqFour$ SYM, the cut in the $s$ channel of the one-loop four-point
amplitude.  We use \eqn{MHVpureg} to move the two negative helicities,
so that they are in locations 1 and 2; {\it i.e.}~we consider
$A_4^\oneloop(1^-,2^-,3^+,4^+)$.  Its $s$-channel cut is depicted
in \fig{RR1lFigure}(a), and is given by,
\bea
C_s &\equiv& {\rm Disc}_s \, A_4^\oneloop(1^-,2^-,3^+,4^+)
\label{scut1lA}\\
&=& \sum_{P\,\in\,\NeqFour} \int{\rm dLIPS}(\ell_1,\ell_3) \,  
A_4^\tree(1^-,2^-,\ell_{3,P},-\ell_{1,P}) \,
A_4^\tree(\ell_{1,P},-\ell_{3,P},3^+,4^+) \,,
\nonumber
\eea
where ${\rm dLIPS}(\ell_1,\ell_3)$ stands for the intermediate
phase-space measure.
In principle, we have to sum over all 16 states $P$ in the multiplet. 
However, the SWI~(\ref{vanishSWI}) and (\ref{vanishmatterSWI}) imply
that there is only one nonvanishing configuration, the one in which
two identical-helicity gluons cross the cut,
\be
C_s = \int {\rm dLIPS}(\ell_1,\ell_3) \,
A_4^\tree(1^-,2^-,\ell_3^+,-\ell_{1}^+)
A_4^\tree(\ell_1^-,-\ell_3^-,3^+,4^+) \,.
\label{scut1lB}
\ee
%

%%%%%%%%% FIGURE %%%%%%%%%%%%%%%
%%%%%%%%%%%%%%%%%%%%

Using \eqn{spabsqrts}, the four-point tree amplitude~(\ref{MHVPT})
can be written as
\be
A_4^\tree(1^-,2^-,3^+,4^+) = 
{ {\spa1.2}^4 \over \spa1.2\spa2.3\spa3.4\spa4.1 } 
= \hbox{phase} \times {s_{12}\over s_{23}} \,.
\label{MHV4simple}
\ee
Substituting for $1/s_{23}$ the appropriate
kinematic variable inside the cut, \eqn{scut1lB} becomes
\bea
C_s &=& \hbox{phase} \times \int {\rm dLIPS}(\ell_1,\ell_3) \,
 { s_{12} \over (\ell_1-k_1)^2 }
 { s_{12} \over (\ell_3-k_3)^2 } 
\nonumber\\
&=& i \, s_{12} s_{23} \, A_4^\tree(1^-,2^-,3^+,4^+)
\int {\rm dLIPS}(\ell_1,\ell_3) \, 
{ 1 \over (\ell_1-k_1)^2 \, (\ell_3-k_3)^2 } \,.
\label{scut1lC}
\eea
This result is represented diagramatically in \fig{RR1lFigure}(a).
The last factor in \eqn{scut1lC} is simply the $s$-channel
cut of the scalar one-loop box integral,
\be
I_4^D(s_{12},s_{23}) = \int {d^D\ell_1 \over (2\pi)^D}
{1 \over \ell_1^2 \, (\ell_1-k_1)^2 \, \ell_3^2 \, (\ell_3-k_3)^2} \,,
\label{box1l}
\ee
because the two propagators with momenta $\ell_1$ and
$\ell_3=\ell_1-k_1-k_2$ are replaced by delta functions in the cut.
Thus an expression for the one-loop four-point amplitude
that matches the $s$-channel cut is,
\be
A_4^{\oneloop}(1^-,2^-,3^+,4^+) =
i \, s_{12} s_{23} \, A_4^\tree(1^-,2^-,3^+,4^+) 
\, I_4^D(s_{12},s_{23}) \,.
\label{final1l}
\ee
Because of \eqn{MHVpureg}, the ratio $A_4^{\oneloop}/A_4^\tree$
is independent of the location of the negative helicities. So we know
that \eqn{final1l} must also match the $t$-channel cut.  Here we only
computed the cut in four dimensions, using helicity states to perform the
intermediate state sum.  However, the same expression~(\ref{final1l})
matches the full $D$-dimensional cuts (when a supersymmetric regulator is
used~\cite{DimRed}), and therefore it must be the correct
answer for the full amplitude, including the dispersive
part~\cite{Green1982sw}.

Furthermore, the two-particle cut can be iterated to higher
loops~\cite{BRY}.  In the next iteration, there is an additional
numerator factor of $s_{12}$ from sewing on the next tree,
while the factor of $1/s_{23}$ becomes an additional propagator.
This result has been abstracted to give the {\it rung rule}~\cite{BRY}
shown in \fig{RR1lFigure}(b), which generates numerator factors for
certain contributions to the $(L+1)$-loop amplitudes, in terms of those
at $L$ loops. Whenever a rung is sewn on perpendicular to two lines 
carrying loop momenta $\ell_1$ and $\ell_2$, an extra factor of
$i(\ell_1+\ell_2)^2$ is generated in the numerator.  An example of
a graph with iterated two-particle cuts, whose numerator can be 
determined by the rung rule, is given in \fig{RR1lFigure}(c).

%%%%%%%%%%%%%%%%%%%%%%%%%%%%%%%%%%%%%%%%%%%%%

\section{KLT copying and rung-rule behavior}
\label{KLTCopyingSection}

Suppose that we have a representation of the $L$-loop
four-point amplitude in $\NeqFour$ SYM, and we want to compute
the $L$-loop four-point amplitude in $\NeqEight$ supergravity.
The KLT relations give us an efficient way of exporting
the information from the first theory to the second.  An
arbitrary generalized cut in $\NeqEight$ supergravity is
given in terms of $\NeqEight$ supergravity tree amplitudes, summed over
all intermediate states.  We rewrite each tree using KLT
in terms of two copies of $\NeqFour$ SYM trees.  The net result is a sum
over products of {\it two copies} of the $\NeqFour$ SYM cuts.
The sum over intermediate states in $\NeqEight$ supergravity
is automatically carried out as a double sum over the $\NeqFour$ SYM states, 
$\sum_{\NeqEight} = \sum_{\NeqFour} \sum_{\NeqFour}$.
Because the KLT relations contain different cyclic orderings 
of the $\NeqFour$ SYM tree amplitudes, we need both planar
(leading-in-$N_c$) and non-planar (subleading-in-$N_c$) terms in
the $L$-loop $\NeqFour$ SYM amplitude.  This is not a surprise; the
gravitational amplitude has no notion of color ordering.

%%%%%%%%% FIGURE %%%%%%%%%%%%%%%
\begin{figure}[t]
\centerline{\epsfxsize 5.2 truein \epsfbox{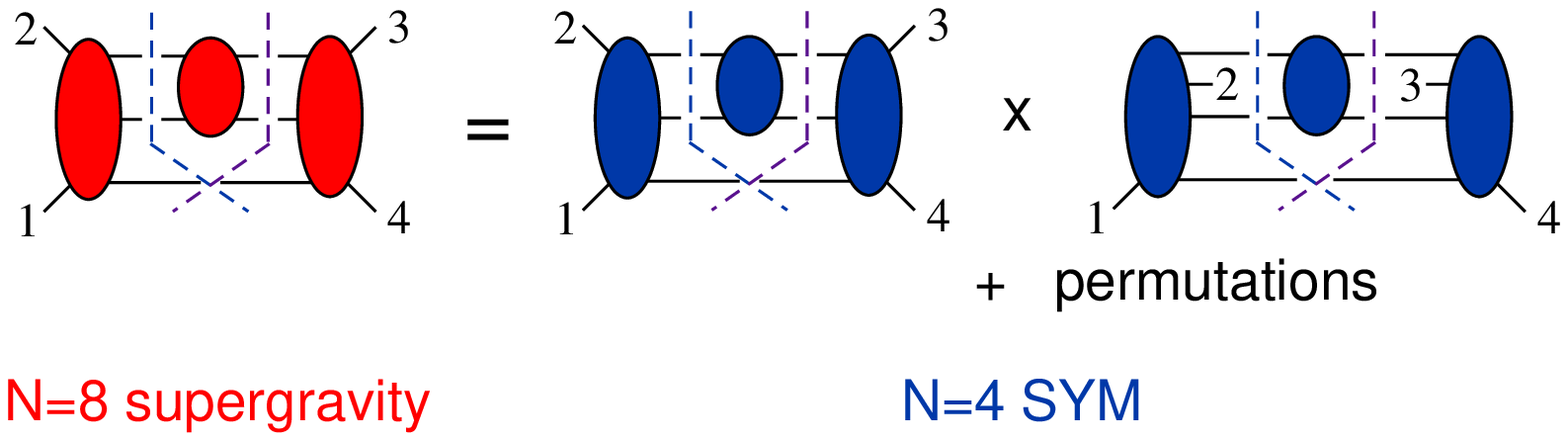}}
\caption[a]{\small Evaluation of a generalized cut in $\NeqEight$
  supergravity at three loops, in terms of planar and non-planar
  cuts in $\NeqFour$ SYM.}
\label{KLTCopyingFigure}
\end{figure}
%%%%%%%%%%%%%%%%%%%%

\Fig{KLTCopyingFigure} shows an example of KLT copying at three loops.
The $\NeqEight$ supergravity cut contains one four-point tree amplitude
and two five-point ones.  We use \eqns{KLTFour}{KLTFive}.
It is convenient to rewrite them as
\bea
M_4^\tree(\ell_1,\ell_2,\ell_3,\ell_4) &=&  
     - i {s_{\ell_1\ell_2}s_{\ell_2\ell_3}\over s_{\ell_1\ell_3}}
  \, A_4^\tree(\ell_1,\ell_2,\ell_3,\ell_4)
  \, \tilde{A}_4^\tree(\ell_1,\ell_2,\ell_3,\ell_4)\,, 
\label{KLTNew} \\
M_5^\tree(1,2,\ell_2,\ell_1,\ell_5)
 &=& -i s_{\ell_5 1} s_{2\ell_2} \, A_5^\tree(1,2,\ell_2,\ell_1,\ell_5)
\, \tilde{A}_5^\tree(1,\ell_1,2,\ell_2,\ell_5)
\, + \, \CP(1,2) \,,
\nonumber\\
M_5^\tree(4,3,\ell_3,\ell_4,5)
 &=& -i s_{\ell_5 4} s_{3\ell_3} \, A_5^\tree(4,3,\ell_3,\ell_4,\ell_5)
 \, \tilde{A}_5^\tree(4,\ell_4,3,\ell_3,\ell_5)
\, + \, \CP(3,4) \,.
\nonumber
\eea
In this way, both occurrences of the four-point $\NeqFour$ SYM
amplitude carry the same cyclic ordering as the $\NeqEight$ supergravity
one, as shown in the figure.  One of the two five-point amplitudes carries
the same ordering, as shown in the left copy, but the other one is
twisted, leading to the right copy.  A reflection symmetry under
the permutation $(1\lr4,\ 2\lr3)$ is preserved by this representation.
The two-fold permutation sum in $M_5^\tree$ in \eqn{KLTNew} leads
to a four-fold permutation sum in the figure; one must add the
permutations $(1\lr2)$, $(3\lr4)$ and $(1\lr2,\ 3\lr4)$.

KLT copying has a simple consequence for terms that are detected
in the maximal cuts, because of the simple relation between
gravity and gauge three-point amplitudes, \eqn{KLTThree}:
the numerator factors are simply squared.

%%%%%%%%% FIGURE %%%%%%%%%%%%%%%
\begin{figure}[t]
\centerline{\epsfxsize 4.7 truein \epsfbox{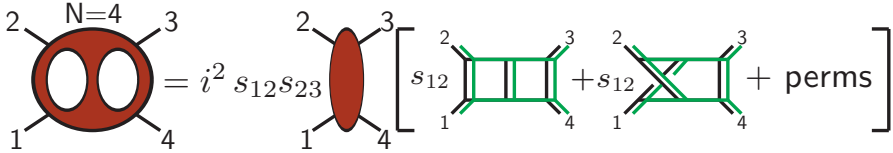}}
\caption[a]{\small The two-loop amplitude in $\NeqFour$ SYM. 
The blob on the right represents the color-ordered tree amplitude 
$A_4^\tree$. In the the brackets, black lines are kinematic $1/p^2$ 
propagators, with scalar ($\phi^3$) vertices. 
Green lines are color $\delta^{ab}$ propagators, with structure 
constant ($f^{abc}$) vertices.}
\label{Neq42loopFigure}
\end{figure}
%%%%%%%%%%%%%%%%%%%%

For example, as mentioned above, the maximal cuts capture the 
full two-loop four-point amplitude in $\NeqFour$ SYM, which
is given by
\bea
A_4^\twoloop &=& - s_{12} s_{23} A_4^\tree \Bigl[
    C^{\P}_{1234} \, s_{12} \, \I_4^{\twoloop,\P}(s_{12}, s_{23}) 
  + C^{\NP}_{1234} \, s_{12}\, \I_4^{\twoloop,\NP}(s_{12},s_{23}) 
\nonumber\\
&& \hskip2.3cm\null
  + \CP(2,3,4) \Bigr]\,,
\label{TwoLoopYM}
\eea
where $\I_4^{\twoloop,(\P,\NP)}$ are the scalar planar and non-planar
double box integrals shown in \fig{Neq42loopFigure}, and
$C^{(\P,\NP)}_{1234}$ are color factors constructed from structure 
constant vertices, with the same graphical structure as the corresponding
integral.  The quantity $s_{12} s_{23} A_4^\tree$ is totally symmetric 
under gluon interchange, and its square is the $R^4$ matrix
element~(\ref{R4matrixelement}), up to a factor of $i$.  
Therefore squaring the prefactors in
\eqn{TwoLoopYM} (and removing the color factors, as appropriate for
gravity) gives the complete two-loop four-point amplitude in 
$\NeqEight$ supergravity,
\bea
M_4^\twoloop &=& -i ( s_{12} s_{23} A_4^\tree )^2 \Bigl[
    s_{12}^2 \, \I_4^{\twoloop,\P}(s_{12}, s_{23}) 
  + s_{12}^2 \, \I_4^{\twoloop,\NP}(s_{12},s_{23}) 
  + \CP(2,3,4) \Bigr] \nonumber\\
 &=& s_{12} s_{23} s_{13} M_4^\tree \Bigl[
    s_{12}^2 \, \I_4^{\twoloop,\P}(s_{12}, s_{23}) 
  + s_{12}^2 \, \I_4^{\twoloop,\NP}(s_{12},s_{23}) 
  + \CP(2,3,4) \Bigr]  \,. \nonumber\\
~{} \label{TwoLoopGr}
\eea
Because the loop integrals appearing in the two amplitudes,
\eqns{TwoLoopYM}{TwoLoopGr}, are precisely the same, the critical 
dimension $D_c$ is automatically the same for both theories at two loops.
This value is $D_c=7$, the dimension in which the two-loop,
seven-propagator integrals, $\sim \, \int d^{2D}\ell/(\ell^2)^7$,
are log divergent.
As mentioned in \sect{DivQGSection}, the divergence is associated with
a counterterm of the form ${\cal D}^4R^4$ in $D=7$.

Another class of diagrams detectable by maximal cuts are the $L$-loop
ladder diagrams.  Their numerator factors are given by the rung rule,
\bea
&&st A_4^\tree \times s^{L-1}
\hskip2.0cm (\hbox{$\NeqFour$ SYM}), \label{Neq4Ladder}\\
&&stu M_4^\tree \times s^{2(L-1)}
\hskip1.25cm (\hbox{$\NeqEight$ supergravity}), \label{Neq8Ladder}
\eea
where the second result was obtained by squaring the first one.
The extra factor of $s$ per loop for gravity corresponds, in the
semi-classical high-energy limit $s \gg t$, to the fact that the
gravitational ``charge'' is the energy in the center of mass,
and it appears squared for each additional rung exchange, That is,
$g^2$ in gauge theory is replaced by
$\kappa^2 E_{\rm CM}^2 = \kappa^2 s$ in gravity.

%%%%%%%%% FIGURE %%%%%%%%%%%%%%%
\begin{figure}[t]
\centerline{\epsfxsize 3.0 truein \epsfbox{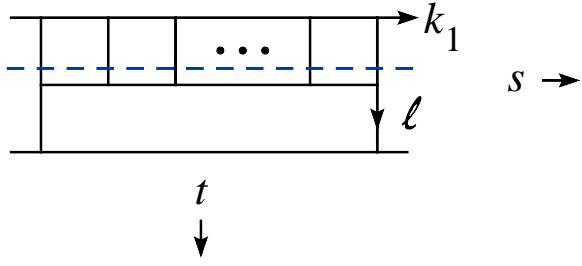}}
\caption[a]{\small Example of a rung-rule contribution with
worse ultraviolet behavior for $\NeqEight$ supergravity than for 
$\NeqFour$ SYM.  The dashed line shows a unitarity cut which
exposes a one-loop $(L+2)$-point amplitude that violates
the no-triangle property.}
\label{UVbadFigure}
\end{figure}
%%%%%%%%%%%%%%%%%%%%

If one now takes an $(L-1)$-loop ladder diagram, and sews two
opposite ends together, one gets the $L$-loop graph shown in
\fig{UVbadFigure}.  The respective rung-rule numerator factors
are now
\bea
&&st A_4^\tree \times t \times [(\ell+k_1)^2]^{L-2}
\hskip2.2cm (\hbox{$\NeqFour$ SYM}), \label{Neq4bad}\\
&&stu M_4^\tree \times t^2 \times [(\ell+k_1)^2]^{2(L-2)}
\hskip1.25cm (\hbox{$\NeqEight$ supergravity}). \label{Neq8bad}
\eea
The extra factors of the external momentum invariant $s$ in the
ladder graph have become factors of $(\ell+k_1)^2$, where $\ell$
is the loop momentum shown in the figure.  This particular contribution,
with $3L+1$ propagators, behaves in the ultraviolet as
\be
\I\ \sim\ \int d^{DL}\ell\ { (\ell^2)^{p(L-2)} \over (\ell^2)^{3L+1} } \,,
\label{UVbadpowercount}
\ee
which is worse for $\NeqEight$ supergravity ($p=2$)
than for $\NeqFour$ SYM ($p=1$).  The critical dimension for
this integral satisfies
$D_C(L) \times L/2 + p(L-2) = 3L+1$.  For $p=1$, this gives the standard
answer for $\NeqFour$ SYM, $D_c(L)=4+6/L$, \eqn{NewPowerCount}.
For $p=2$, if there are no additional cancellations, 
it gives $D_c(L)=2+10/L$, \eqn{OldPowerCount}, suggesting a
divergence in $D=4$ at $L=5$.

However, there were reasons
to expect additional cancellations beginning at $L=3$~\cite{Finite},
which were revealed by a particular cut of \fig{UVbadFigure},
shown as the dashed line slicing through all the rungs of the ladder.
This $L$-particle
cut has a one-loop $(L+2)$-point amplitude on one side of it with
$4(L-2)$ factors of $L$ in the numerator.  This one-loop integral would
have generated, after reduction to a basis of scalar integrals,
nonvanishing coefficients for triangle integrals, in contradiction with
the no-triangle hypothesis~\cite{NoTriangleB}. This hypothesis was a
known fact~\cite{OneloopMHVGravity} already for the five-point amplitude
encountered at $L=3$, and would later be proved for an arbitrary number of
legs~\cite{NoTriangleProof,AHCKGravity}.  Therefore other graphs had
to combine to cancel its large $\ell$ behavior, integrals that were
not detectable in the two-particle cuts.  This result inspired the
computation of the complete four-graviton amplitude in $\NeqEight$
supergravity at three, and later four, loops.

%%%%%%%%%%%%%%%%%%%%%%%%%%%%%%%%%%%%%%%%%%

\section{$\NeqEight$ supergravity at three loops}
\label{Neq8ThreeLoopSection}

%%%%%%%%% FIGURE %%%%%%%%%%%%%%%
\begin{figure}[t]
\centerline{\epsfxsize 5.5 truein \epsfbox{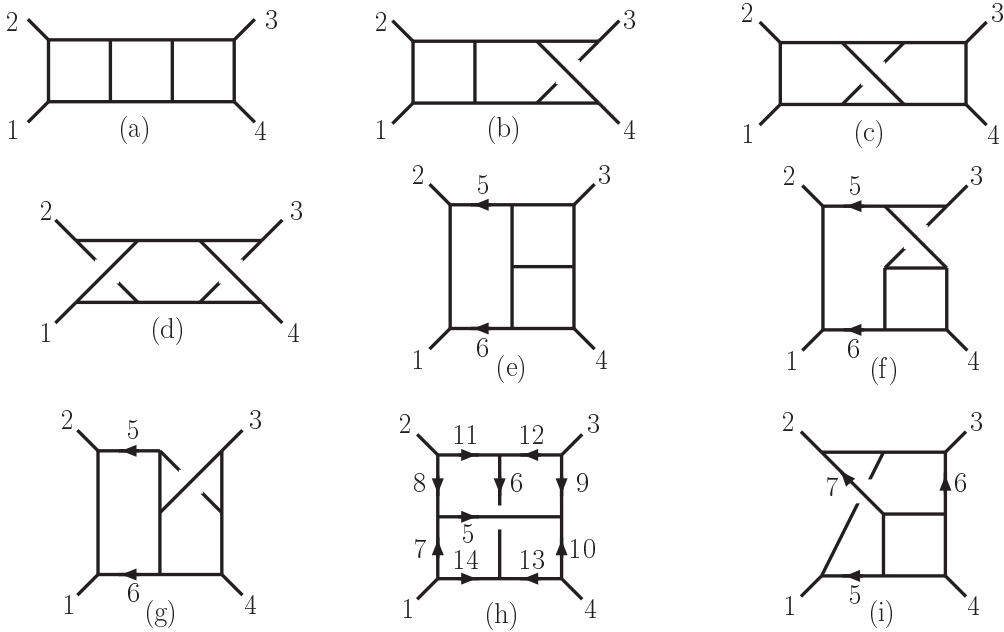}}
\caption[a]{\small The different cubic four-point graphs in terms of which 
four-point three-loop amplitudes may be expressed.}
\label{IntegralsThreeLoopFigure}
\end{figure} 
%%%%%%%%%%%%%%%%%%%%%%%%%%%%%%%%

\Fig{IntegralsThreeLoopFigure} shows the nine cubic four-point graphs
at three loops that do not contain three-point subdiagrams.
Both the $\NeqFour$ SYM and $\NeqEight$ supergravity amplitudes
can be described by giving the loop-momentum numerator polynomials $N^{(x)}$
for these graphs.  (A recent, alternate representation~\cite{BCJ10}
makes use of the graphs with three-point subdiagrams, but they are not
necessary.)  In addition, the $\NeqFour$ SYM graphs are to be multiplied
by the corresponding color structure, as in \fig{Neq42loopFigure}.
Notice that seven of the nine graphs, (a)--(g), have iterated two-particle
cuts, so they can be computed via the rung rule.  Only (h) and (i) require
additional work.

%%%%%%%%%%%% TABLE %%%%%%%%%%%%%%%%%%%%%%%%%%
\begin{table*}
\caption{ \small
The numerator factors $N^{(x)}$ for the integrals $I^{(x)}$ in
\fig{IntegralsThreeLoopFigure} for $\NeqFour$ super-Yang-Mills theory. 
The first column labels the integral, the second
column the relative numerator factor. An overall factor of 
$s_{12} s_{23} A_4^\tree$ has been removed.  The invariants
$s_{ij}$ and $\tau_{ij}$ are defined in \eqn{InvariantsDef}.
\label{NumeratorYMTable} }
\vskip .4 cm
\begin{center}
\begin{tabular}{||c|c|}
\hline
Integral $I^{(x)}$ & $N^{(x)}$ for $\NeqFour$ super-Yang-Mills \\
\hline
\hline
(a)--(d) &  $
%%%%% begin : NYMad
s_{12}^2
%%%%% end : NYMad
$ 
  \\
\hline
(e)--(g) &  $
%%%%% begin : NYMeg
s_{12} \,s_{46}
%%%%% end : NYMeg
$  \\
\hline
(h)&  $\; 
%%%%% begin : NYMh
s_{12} (\tau_{26} + \tau_{36}) +
       s_{23} (\tau_{15} + \tau_{25}) + 
       s_{12} s_{23} 
%%%%% end : NYMh
\;$
\\
\hline
(i) & $\;
%%%%% begin : NYMi
s_{12} s_{45} - s_{23} s_{46} -  {1\over 3} (s_{12} - s_{23}) \ell_7^2
%%%%% end : NYMi
 \;$  \\
\hline
\end{tabular}
\end{center}
\end{table*}
%%%%%%%%%%%%%%%%%%%%%%%%%%%%%%%%%

\Tab{NumeratorYMTable} gives the values of $N^{(x)}$ for $\NeqFour$ SYM
in terms of the following invariants,
\bea
&& s_{ij} = (k_i +k_j)^2 \,,    \hskip 5.0 cm (i,j \le 4) \nn \\
&& s_{ij} = (k_i +\ell_j)^2 \,,    \hskip 1.5 cm 
\tau_{ij} = 2 k_i\cdot \ell_j\,,   \hskip 1.1 cm (i \le 4,\ \ j\ge5) \nn \\
&& s_{ij} = (\ell_i + \ell_j)^2 \,.    \hskip 5.1 cm  (i, j\ge5) 
\label{InvariantsDef}
\eea
Note that $s_{ij}$ is quadratic in the loop momenta $\ell_j$,
if $j>4$, but $\tau_{ij}$ is linear.  Every $N^{(x)}$ is manifestly
quadratic in the loop momenta, consistent with $D_c(L)=4+6/L$ for $L=3$.
It is easy to verify the numerators for graphs (a)--(g) using the rung
rule. 

%%%%%%%%%%%% TABLE %%%%%%%%%%%%%%%%%%%%%%%%%%
\begin{table*}
\caption{ \small
Numerator factors $N^{(x)}$ for $\NeqEight$ supergravity.
The first column labels the integral, the second column the relative
numerator factor. An overall factor of 
$s_{12} s_{13} s_{14} M_4^\tree$ has been removed.
\label{NumeratorGravityTable} }
\vskip .4 cm
\begin{center}
\begin{tabular}{||c|c||}
\hline
Integral $I^{(x)}$ & $N^{(x)}$ for $\NeqEight$ supergravity  \\
\hline
\hline
(a)--(d) 
& $
\vphantom{\Bigr|}
%%%%% begin : NGravad
 [s_{12}^2]^2
%%%%% end : NGravad
$ \\
\hline
(e)--(g) & $\vphantom{\Bigr|}
%%%%% begin : NGraveg
 s_{12}^2 \, \tau_{35} \, \tau_{46}
%%%%% end : NGraveg
$  \\
\hline
(h) & $\;
%%%%% begin : NGravhA
(s_{12} (\tau_{26} + \tau_{36}) + s_{23}(\tau_{15}+\tau_{25})+s_{12} s_{23})^2
%%%%% end : NGravhA
\;$ \\
& $\; \null
%%%%% begin : NGravhB
   + (s_{12}^2 (\tau_{26} + \tau_{36})
   -  s_{23}^2 (\tau_{15} + \tau_{25})) 
             (\tau_{17} + \tau_{28} + \tau_{39} + \tau_{4,10})
%%%%% end : NGravhB
         \; $  \\
    &  $ \; \null
%%%%% begin : NGravhC
      + s_{12}^2 (\tau_{17} \tau_{28} + \tau_{39} \tau_{4,10})
      + s_{23}^2 (\tau_{28} \tau_{39} + \tau_{17} \tau_{4,10} )
           + s_{13}^2  (\tau_{17} \tau_{39} + \tau_{28} \tau_{4,10}) 
%%%%% end : NGravhC
\;     \vphantom{\Bigl( \Bigr)_{A_A} }
         $   \\
\hline
(i) & $
\vphantom{\bigl|_{A_A}} 
%%%%% begin : NGraviA
(s_{12}\, \tau_{45} - s_{23} \, \tau_{46})^2 
- \tau_{27} (s_{12}^2 \,\tau_{45} + s_{23}^2 \,\tau_{46})
- \tau_{15} (s_{12}^2 \,\tau_{47} + s_{13}^2 \,\tau_{46})
%%%%% end : NGraviA
 \; $ \\
& $ \; \null
%%%%% begin : NGraviB
- \tau_{36} (s_{23}^2\, \tau_{47} + s_{13}^2 \,\tau_{45})
+ \ell_5^2 \, s_{12}^2 \,s_{23} 
+ \ell_6^2 \, s_{12} \, s_{23}^2 
- {1\over 3} \ell_7^2 \, s_{12} \, s_{13} \, s_{23}
%%%%% end : NGraviB
\; $ \\
\hline
\end{tabular}
\end{center}
\end{table*}
%%%%%%%%%%%%%%%%%%%%%%%%%%%%%%%%%%%%%

\Tab{NumeratorGravityTable} gives the values of $N^{(x)}$ for
$\NeqEight$ supergravity, in a form~\cite{CompactThree} which
is also manifestly quadratic in the loop momenta. (In an earlier version
of the amplitude~\cite{GravityThree}, the quadratic nature was not yet
manifest.)  Comparing the two sets of numerators, we see that the 
$\NeqEight$ supergravity ones are the squares of the $\NeqFour$ SYM
ones, up to contact terms, as expected from the KLT relations.
For example, in graphs (e)--(g), 
$s_{46} = \tau_{46} + \ell_6^2 = \tau_{35} + \ell_5^2$,
so $s_{12}^2 \tau_{35} \tau_{46} \approx [s_{12} s_{46}]^2$ (modulo
$\ell_i^2$ terms).

Because the numerator factors for both $\NeqEight$ supergravity and
$\NeqFour$ SYM are manifestly quadratic in the loop momenta,
the critical dimension $D_c$ at three loops should continue to obey
\eqn{NewPowerCount}, {\it i.e.}~$D_c=6$.  Indeed, when the ultraviolet
poles in the nine integrals for $\NeqEight$ supergravity are evaluated, no
further cancellation is found, and the resulting pole is
\be
M_4^{(3),D=6-2\e} \bigr|_{\rm pole}
\, = \, \frac{1}{\epsilon}\,{5 \, \zeta_3\over (4\pi)^9} \,
(s_{12} s_{23} s_{13})^2 M_4^\tree  \,, \hskip .3 cm 
\label{ThreeLoopD6Div}
\ee
corresponding to a counterterm of the form ${\cal D}^6R^4$ in $D=6$.
(The form of this divergence was recently reproduced from duality
arguments in string theory~\cite{GRV2010}; however, the rational number
does not agree with \eqn{ThreeLoopD6Div} as of this writing.
Whether or not this indicates an issue in decoupling massive
states~\cite{GOS} remains unclear.)

%%%%%%%%%%%%%%%%%%%%%%%%%%%%%%%%%%%%%%%%%%

\section{$\NeqEight$ supergravity at four loops}
\label{Neq8FourLoopSection}

\begin{figure}[t]
\centerline{\epsfxsize 5.5 truein \epsfbox{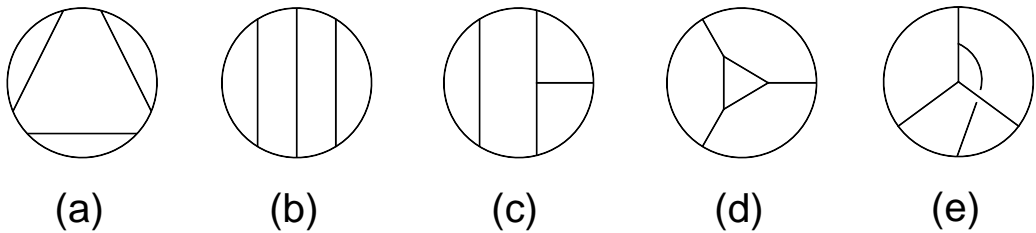}}
\caption[a]{\small Cubic vacuum graphs at four loops.}
\label{vac4Figure}
\end{figure}

The general strategy~\cite{Neq84,FourLoopYM} used to compute
the four-loop four-point amplitude in $\NeqEight$ supergravity is the same
as at three loops; however, the bookkeeping issues are considerably
greater.  One can start by classifying the cubic vacuum graphs.
At three loops there were only two; at four loops there are five,
shown in \fig{vac4Figure}.  

The next step is to decorate the five vacuum graphs with four external
legs to get the cubic four-point graphs.  As at lower loops, graphs
containing triangles (three propagators or fewer on a loop) or other
three point subgraphs can be dropped.  \Fig{vac4Figure}(a) only gives rise
to triangle-containing graphs, so it can be neglected henceforth.
Altogether there are 50 cubic four-point graphs with nonvanishing numerators.
Graphs (b) and (c) do generate no-triangle four-point graphs, but the
numerators for all such graphs can be determined, up to possible
contact terms, using the rung rule.  For this reason,
their associated numerator polynomials are the simplest.
Graphs (d), and particularly (e), give rise to the most complex
numerators.

%%%%%%%%% FIGURE %%%%%%%%%%%%%%%
%%%%%%%%%%%%%%%%%%%%%%%%%%%%%%%%

\begin{figure}[t]
\centerline{\epsfxsize 5.8 truein \epsfbox{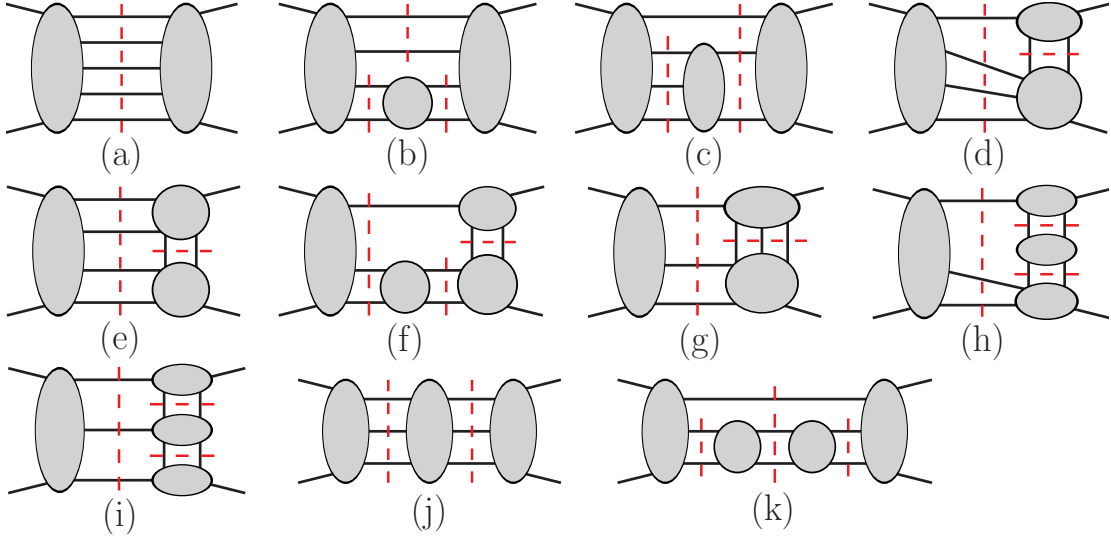}}
\caption[a]{\small Generalized cuts which, along with iterated
two-particle cuts, suffice to determine any massless four-point
four-loop amplitude.}
\label{CutBasisFigure}
\end{figure}

The numerators are first determined for $\NeqFour$ SYM using
the method of maximal cuts.  At four loops the maximal cuts have 13 cut 
conditions $\ell_i^2=0$.  Then near-maximal cuts with only 12 cut conditions 
are considered, followed by ones with 11.  At this point the 
$\NeqFour$ SYM ansatz is complete; no more terms need to be added.
The result is verified by comparison against many of the generalized
cuts with real momenta, shown in \fig{CutBasisFigure}.  Cases (a) and (i)
involve six- and seven-point next-to-MHV amplitudes, for which the sums
over super-multiplets crossing the cuts are more intricate than when all
amplitudes are MHV.  These cuts were evaluated using
compact results for the super-sums obtained 
by Elvang, Freedman and Kiermaier~\cite{EFK08}.

%%%%%%%%% FIGURE %%%%%%%%%%%%%%%%%%
%%%%%%%%%%%%%%%%%%%%%%%%%%%%%%%%

The 50 numerator polynomials for $\NeqEight$ supergravity are then
constructed using information provided by the KLT relations. The results
are quite lengthy, but are provided as {\sc Mathematica} readable files in
ref.~\cite{Neq84}, along with some tools for manipulating them.

From the numerator polynomials, the ultraviolet behavior of the amplitude
can be extracted.  One has to expand the integrals in the limit of small
external momenta, relative to the loop momenta~\cite{TaylorExpand}.
If the ultraviolet behavior is manifest, as at three loops with
the representation found in ref.~\cite{CompactThree}, then only the first term
in this expansion is required.  At four loops, it was necessary to go to
third order to see all observed cancellations.  More concretely, 
the numerator polynomials, omitting an overall factor of $stu M_4^\tree$,
have a mass dimension of 12, {\it i.e.}~each term is of the form
$k^{12-m} \ell^m$, where $k$ and $\ell$ stand respectively for external
and loop momenta.  The maximum value of $m$ turns out to be 8, for every
integral.  The integrals all have 13 propagators, so they have the form
$\I \sim \int d^{4D}\ell \, \ell^{8-26}$.  The amplitude is manifestly
finite in $D=4$, because $4\times4+8-26<0$.  (This result is not
unexpected, given the absence of a ${\cal D}^2R^4$
counterterm~\cite{DHHK,Kallosh4loop}.)  The amplitude is not
manifestly finite in $D=5$; to see that requires cancellation of the 
$k^4\ell^8$, $k^5\ell^7$ and $k^6\ell^6$ terms, after expansion around
small $k$.

The cancellation of the $k^4\ell^8$ terms is relatively simple;
one can set the external momenta $k_i$ to zero, and collect the
coefficients of the two resulting vacuum graphs, (d) and (e)
in \fig{vac4Figure}, observing that the (d) and (e) coefficients
both vanish.  The cancellation
of the $k^5\ell^7$ terms (and the $k^7\ell^5$ terms) is trivial:
Using dimensional regularization, with no dimensionful parameter,
Lorentz invariance does not allow an odd-power divergence.
The most intricate cancellation is of the $k^6\ell^6$ terms,
corresponding to the vanishing of the coefficient of the potential
counterterm ${\cal D}^6R^4$ in $D=5$.  In the expansion of the integrals
to the second subleading order as $k_i \to 0$, 30 different four-loop 
vacuum integrals are generated.  However, there are consistency relations
between the integrals, corresponding to the ability to shift the loop
momenta by external momenta before expanding around $k_i = 0$.
These consistency relations are powerful enough to imply the cancellation 
of the ultraviolet pole in $D=5-2\e$.  As a check, we evaluated all the
ultraviolet poles directly, with the same conclusion.

In summary, the four-loop four-point amplitude of $\NeqEight$ 
supergravity is ultraviolet finite for $D < 11/2$~\cite{Neq84}, the same
critical dimension as for $\NeqFour$ super-Yang-Mills theory.
Finiteness in $5\le D<11/2$ is a consequence of nontrivial cancellations, 
beyond those already found at three loops~\cite{GravityThree,CompactThree}. 
These results provide the strongest direct support to date
for the possibility that $\NeqEight$ supergravity might be a
perturbatively finite quantum theory of gravity. 

%%%%%%%%%%%%%%%%%%%%%%%%%%%%%%%%%%%%%%%%%%%%%%%%%%%%

\section{Conclusions and Outlook}
\label{ConclusionsSection}

What are the prospects for going beyond four loops?  As mentioned in the
Introduction, there are indications of a
problem~\cite{GRV2010,Vanhove2010,BG2010}
with the pure spinor formalism~\cite{Berkovits} at five loops, which
could lead to a deviation from the critical dimension
formula~(\ref{NewPowerCount}) at this order.  It would be very interesting
to clarify the situation by computing the complete amplitude.
This would be a major undertaking, but recent developments might provide
some assistance~\cite{BCJ08,BCJ10,BDHK10}.
It is also worth noting that the unitarity method generically implies
sets of higher-loop cancellations as a consequence of lower-loop
ones~\cite{Finite}, by finding the latter contributions as subgraphs
exposed by unitarity cuts.  The fact that certain higher-loop cancellations 
were implied by the no-triangle behavior of one-loop $n$-point
amplitudes was discussed in \sect{KLTCopyingSection}, but the
excellent ultraviolet behavior of the two-, three- and four-loop
four-point amplitudes has similar implications.  Combining this
information properly is still an issue, however. 

The unitarity method works with on-shell objects, so it can maintain
more supersymmetry than is generally possible with off-shell Feynman
diagrams.  However, not all symmetries of $\NeqEight$ supergravity
are kept manifest in the current approach.  In particular the theory
has a continuous noncompact coset symmetry,
$E_{7(7)}$~\cite{CremmerJulia}.  The 70 massless scalars
parametrize the coset space $E_{7(7)}/SU(8)$, and the non-$SU(8)$ part
of the symmetry is realized nonlinearly.  When using the KLT relations, 
there is an $SU(4)$ $R$ symmetry associated with each $\NeqFour$
supersymmetry, so only an $SU(4)\times SU(4)$ subgroup of the 
linearly-realized $SU(8)$ is kept manifest, and none of the
$E_{7(7)}/SU(8)$.
There has been some recent progress on the role of $E_{7(7)}$ in 
$\NeqEight$ supergravity: using it to
construct terms in the light-cone superspace Hamiltonian~\cite{BKR};
describing its explicit action on covariant fields~\cite{KalloshSoroush};
exploring its implications for tree-level $S$ matrix
elements~\cite{BEZ,AHCKGravity,Kallosh2008rr}, and also at 
loop level~\cite{Kallosh2008ru,He2008pb}; and assessing the $E_{7(7)}$
invariance of the $R^4$ counterterm~\cite{BD09}.
However, a deeper understanding of the implications of $E_{7(7)}$
invariance would certainly be welcome.

Suppose that $\NeqEight$ supergravity is finite to all loop orders.
This still would not prove that it is a nonperturbatively consistent
theory of quantum gravity.  There are at least two reasons to think that
it might need a nonperturbative ultraviolet completion:
\begin{enumerate}
\item The (likely) $L!$ or worse growth of the coefficients of the 
order $L$ terms in the perturbative
expansion, which for fixed-angle scattering, means a non-convergent 
behavior $\sim L! \, (s/M_{\rm Pl}^2)^L$.
\item The fact that the perturbative series seems to be $E_{7(7)}$
invariant, while the mass spectrum of black holes is non-invariant (see
{\it e.g.} ref.~\cite{BFK} for recent discussions).
\end{enumerate}
QED is an example of a perturbatively well-defined theory that needs 
an ultraviolet completion;
it also has factorial growth in its perturbative coefficients,
$\sim L! \, \alpha^L$, due to ultraviolet renormalons associated with the
Landau pole.  Of course, for small values of $\alpha$, QED works extremely
well; for example, it predicts the anomalous magnetic moment of the
electron to 10 digits of accuracy.  Also, we know of many pointlike
nonperturbative ultraviolet completions for QED, namely asymptotically free
grand unified theories.  Are there any imaginable pointlike completions
for $\NeqEight$ supergravity?  Maybe the only completion is string theory;
or maybe this cannot happen because of the impossibility of decoupling
nonperturbative states~\cite{GOS}. 

In any event, it is clear that the remarkably good perturbative 
ultraviolet behavior of $\NeqEight$ supergravity has provided many 
surprises to date.  Although the theory may not be of direct
phenomenological interest, perhaps it could some day point the way
to other, more realistic finite theories.  As a ``toy model'' for
a pointlike theory of quantum gravity, it has been extremely instructive,
and is still well worth exploring further.

%%%%%%%%%%%%%%%%%%%%%%%%%%%%%%%%%%%%%%%%%%%

\section*{Acknowledgments}
\vskip -.3 cm 

I am grateful to Zvi Bern, John Joseph Carrasco, Henrik Johansson, David
Kosower and Radu Roiban for collaboration on many of the topics described
here, and to Zvi Bern for comments on the manuscript.
I thank the organizers of the 47$^{\rm th}$ International School of
Subnuclear Physics for the opportunity to present these lectures in such a
stimulating and pleasant environment.  This work was supported by the US
Department of Energy under contract DE-AC02-76SF00515.

\baselineskip 15pt

\parskip=.1cm

%%%%%%%%%%%%%%%%%%%%%%%%%%%%%%%%%%%%%%%%%%%%%%%%%%%%%%%%%%

\end{document}